\documentclass[a4paper,fleqn,usenatbib]{mnras}

\usepackage[T1]{fontenc}
\usepackage{ae,aecompl}
\usepackage[dvipsnames]{xcolor}

\usepackage{graphicx}	
\usepackage{amsmath}	
\usepackage{amssymb}	

\title[3D heat transfer in magnetized NS]{Three-dimensional heat transfer effects in external layers of a magnetized neutron star}

\author[I.A. Kondratyev et al.]{Ilya A. Kondratyev,$^{1,2}$\thanks{E-mail: mrkondratyev95@gmail.com}
	Sergey. G. Moiseenko,$^{1,2}$\thanks{E-mail: moiseenko@iki.rssi.ru} 
	Gennady S. Bisnovatyi-Kogan,$^{1,3,4}$\thanks{E-mail: gkogan@iki.rssi.ru}
	\newauthor and Maria V. Glushikhina$^{1}$\thanks{E-mail: m.glushikhina@iki.rssi.ru}
	\\
	$^{1}$ Space Research Institute RAS, Profsoyuznaya st. 84/32, Moscow, Russia, 117997\\
	$^{2}$ National Research University Higher School of Economics, Dept. of Physics,\\ Staraya Basmannaya st. 21/4 b.5, Moscow, Russia, 105066\\
	$^{3}$ National Research Nuclear University MEPhI, Kashirskoe sh., 31, Moscow, Russia, 115409\\
	$^{4}$ Moscow Institute of Physics and Technology, Institutskiy per. 9, Dologoprudny, Moscow region, Russia, 141701
}

\date{Accepted XXX. Received YYY; in original form ZZZ}

\pubyear{2020}

\begin{document}
	\label{firstpage}
	\pagerange{\pageref{firstpage}--\pageref{lastpage}}
	\maketitle
\begin{abstract}
Determination of a magnetic field structure on a neutron star (NS) surface is an important problem of a
modern astrophysics. In a presence of strong magnetic fields a thermal conductivity of a degenerate matter
is anisotropic. In this paper we present 3D anisotropic heat transfer simulations in outer layers of
magnetized NSs, and construct synthetic thermal light curves. We have used a different from previous works tensorial
thermal conductivity coefficient of electrons, derived from the analytical solution of the Boltzmann
equation by the Chapman-Enskog method. We have obtained a NS surface temperature distribution in
presence of dipole-plus-quadrupole magnetic fields. We consider a case, in which magnetic axes of a dipole
and quadrupole components of the magnetic field are not aligned. To examine observational manifestations
of such fields we have generated thermal light curves for the obtained temperature distributions using
a composite black-body model. It is shown, that the simplest (only zero-order spherical function in
quadrupole component) non-coaxial dipole-plus-quadrupole magnetic field distribution can significantly
affect the thermal light curves, making pulse profiles non-symmetric and amplifying pulsations in
comparison to the pure-dipolar field.
\end{abstract}




\section{Introduction}
A strength of magnetic fields on the NS surface may reach $\sim 10^{12-13}$G, and $10^{15}$G in magnetars. One of possible ways to observe surface
magnetic fields is an observation of a thermal radiation in the soft X-ray band (e.g. \cite{obs1, obs2, obs3}).
X-ray observatories, such as ROSAT, Chandra and XMM-Newton, have detected thermally emitting compact objects. Seven nearby radio-silent XDINSs (X-ray Dim Isolated
Neutron Star) are called $"$magnificent seven$"$ (see e.g. (\cite{turolla09}) for a review). Periodic changes in spectra of such NSs may indicate to non-uniform
temperature distributions on their surfaces. Such heterogeneities are determined by an anisotropic thermal
conductivity of degenerate matter in presence of a strong magnetic field. Outer layers of a NS consist of plasma with degenerate electrons,
and non-degenerate non-relativistic nuclei. The pressure is determined mostly by the electrons, and the matter can form a state of the Coulomb
crystal or liquid, and a heat transfers mostly by the electrons as well. The thermal conductivity is suppressed across the magnetic field
lines.
A degree of its suppression across the magnetic field is determined by the so called magnetization parameter $\omega\tau$, where $\tau$ is the average time between electron-nuclei collisions, where $\omega = eB/m_e^*c$ is the electron cyclotron plasma frequency, with $m_e^* = m_e\sqrt{1 + {p_{fe}}^2 /m_e^2c^2}$ as an effective relativistic electron mass, and $p_{fe} = \hbar(3\pi^2n_e)^{1/3}$ as an electron Fermi
momentum, $e$ as the electron charge, $c$ as the speed of light, $\hbar$ as the reduced Planck constant. A
thermal conductivity tensor, as well as other kinetic coefficients, for the plasma with degenerate electrons
are
derived in a series of papers by \cite{bisn1,bisn2}, \cite{glu20}, from the solution of the Boltzmann equation by the
Chapman-Enskog method. The degree of the heat flux depression across the field is stronger, than in previous
works(e.g. \cite{fi76, YakUrp}), where a ratio between heat conductivities along and across the magnetic field lines is connected by the relation
$\frac{\kappa_\parallel}{\kappa_\perp} = 1 + (\omega\tau)^2$.

Heat transfer processes in the outer layers of the NS with the coaxial dipolar and quadrupolar fields were
considered earlier in the paper \cite{kondmoisBKGl}, hereinafter referred as Paper I, where the model of the
magnetized envelope and used numerical technique are described in details.
In this paper we obtain stationary temperature distribution in outer layers of NSs in the presence of dipole
and
quadrupole fields, whose axes are not aligned. We solved numerically a three dimensional heat transfer equation
in the NS crust for the densities $\rho = 10^{10}-2\cdot10^{14}$ g/cm$^3$ using our $T_s-T_b$- relationship for
the outer magnetized envelope, which connects temperature $T_b$ on the $\rho = 10^{10}$ g/cm$^3$ with the
temperature $T_s$ on the NS surface, similarly to \cite{G1983, PY01,kondmoisBKGl}. We have built $"T_s-T_b"$-relationship
adopting thermal conductivities from \cite{bisn1} in Paper 1.
In a 2D case a problem about finding a stationary solution of the NS temperature is was studied by several
authors \cite{heat1, heat2, heat3}, and a 2D NS cooling problem was considered by \cite{heat4, heat5, heat6}
(see also a review by \cite{pot2015}). In this work we restrict ourselves to consideration of stationary
temperature distributions. We mention also a recent review \cite{ponsvig2019} on numerical simulations of the magnetic field and thermal evolution of an isolated NS.

The paper is organized as follows. In second part of this work we review a basic physical input, such as
properties of heat transfer, magnetic field configurations and equation of state. In the third part
we briefly discuss a thermal structure of a magnetized outer envelope. In the forth part a formulation of the boundary problem for the 3D heat transfer equation is given. In the fifth part we present the results about the
temperature distributions, and thermal light curves. In appendices we discuss radiative opacities, and numerical
algorithm used in the calculations.

\section{Physical input}
\subsection{Heat transfer in presence of a magnetic field}
The temperature distribution is determined by the heat transfer equation
\begin{equation}
C\frac{\partial T }{\partial t} = \nabla\cdot \boldsymbol{\hat{\kappa}}\cdot\nabla T + f
\end{equation}
where $C$ is a heat capacity, $\boldsymbol{\hat{\kappa}}$ is a thermal conductivity tensor, $f$ is defined by
heat sources and sinks (Joule heating, neutrino emission, etc.). We look for a stationary solution
($\frac{\partial T }{\partial t} = 0$) in absence of sources and sinks, with $f=0$.

The thermal conductivity tensor $\boldsymbol{\hat{\kappa}}$ for strongly degenerate electrons in the magnetic field was
obtained by~\cite{bisn1} using the Chapman-Enskog method for the Boltzmann equation. This tensor takes into
account heat fluxes along and across the magnetic field as well as the Hall heat flux. In the Cartesian
coordinates it is written as follows
\begin{equation}
\begin{cases}
{\kappa}_{ij} = \frac{k_B^2Tn_e}{m_e^*}\tau\big(\kappa^{(1)}\delta_{ij} +
\kappa^{(2)}\varepsilon_{ijk}\frac{B_k}{B} + \kappa^{(3)}\frac{B_iB_j}{B^2}\big) \\
\kappa^{(1)} = \frac{5\pi^2}{6}\big(\frac{1}{1 + (\omega\tau)^2} - \frac{6}{5}\frac{(\omega\tau)^2}{(1 +
(\omega\tau)^2)^2} \big) \\
\kappa^{(2)} = -\frac{4\pi^2}{3}\omega\tau\big(\frac{1}{1 + (\omega\tau)^2} -
\frac{3}{4}\frac{(\omega\tau)^2}{(1 + (\omega\tau)^2)^2} \big) \\
\kappa^{(3)} = \frac{5\pi^2}{6}(\omega\tau)^2\big(\frac{1}{1 + (\omega\tau)^2} + \frac{6}{5}\frac{1}{(1 +
(\omega\tau)^2)^2} \big)
\end{cases}
\label{heatcond}
\end{equation}
where $n_e = \frac{\rho Z}{Am_u}$ is an electron number density, $Z$ is the nucleus (ion) charge number and $A$
is the mass number; $\tau = \frac{3}{32\pi^2}\frac{h^3}{m_e^*Ze^4\Lambda}$ is an average time between
electron-nuclei collisions, $k_B$ is the Boltzmann constant, $h$ is the Planck constant, $\Lambda$ is the
Coulomb logarithm (see e.g. \cite{bragsi}), we adopt its value from the paper by \cite{YakUrpL}. Parameter
$\omega\tau$ changes drastically in the crust and the envelope of the NS. Its value is close to unity at $\rho \sim 10^{10}$ g/cm$^3$, $B \sim 10^{13}$G, and approximately is changing  $\omega\tau \sim B/\rho^{2/3}$ for the ultra-relativistic degenerate electron gas in the crust.
 As it follows from \eqref{heatcond}, the heat conductivity coefficients across and along the magnetic field may be
written as following

\begin{equation}
\begin{cases}
\kappa_{e\perp} = \frac{k_B^2Tn_e}{m_e^*}\tau\kappa^{(1)},\\
\kappa_{e\parallel} = \frac{k_B^2Tn_e}{m_e^*}\tau\big(\kappa^{(1)} + \kappa^{(3)}\big).
\end{cases}
\label{heatfl}
\end{equation}

Note here one important detail. The value of a thermal conductivity of a strongly degenerate electron gas along the magnetic field $\kappa'_{e\parallel}$, from papers by \cite{fi76,YakUrp}, has a different numerical coefficient from\eqref{heatfl} for $\kappa_{e\parallel}$:
\begin{equation}
\kappa'_{e\parallel} =\frac{\pi^2}{3}\frac{k_B^2Tn_e}{m_e^*}\tau,
\label{heatcondterr}
\end{equation}
which is 2.5 times less than in \eqref{heatcond}. In a large amount of astrophysical studies electron thermal conductivity coefficient is used in that setting. Expression \eqref{heatcondterr} is used for the thermal electron conductivity in metals in laboratory, in the condition of zero electron diffusion velocity (electrical current).

This discrepancy can be shown clear for the thermal conduction coefficient in the absence of a magnetic field, as well as the one
along the magnetic field. As it follows from the Boltzmann equation in the Lorentz approximation, the heat flux  ${\bf q}$, and the average electron velocity $\langle{\bf v_e}\rangle$ are defined as (e.g. \cite{bisn2001})
\begin{equation}
\begin{gathered}
{\bf q} = -\frac{640k_B}{\Lambda}\frac{m_e(k_BT)^4}{n_NZ^2e^4h^3}(G_5 -
\frac{1}{2}\frac{G_{5/2}}{G_{3/2}}G_4)\cdot\nabla T - \\
\frac{128}{\Lambda}\frac{m_e(k_BT)^5}{n_NZ^2e^4h^3}\frac{G_{5/2}}{G_{3/2}}G_4\cdot{\bf d_e} \\
\langle{\bf v_e}\rangle = -\frac{128k_B}{\Lambda}\frac{m_e(k_BT)^3}{n_Nn_eZ^2e^4h^3}(G_4 -
\frac{5}{8}\frac{G_{5/2}}{G_{3/2}}G_3)\cdot\nabla T -\\
\frac{32}{\Lambda}\frac{m_e(k_BT)^4}{n_Nn_eZ^2e^4h^3}\frac{G_{5/2}}{G_{3/2}}G_3\cdot{\bf d_e},
\end{gathered}
\label{boltzmannTC}
\end{equation}
where $n_N$ is a nuclei number density, a vector ${\bf d_e}$ determines a diffusive flux (see \cite{bisn2}).
A function $G_n = G_n(x_0)$ is the Fermi function, $x_0 = \frac{p_{fe}^2}{2m_ek_BT}$. 

In laboratory conditions, when the electrical conductivity is small, and electrical currents are damped rapidly, so that the current density ${\bf j_e} \sim \langle{\bf v_e}\rangle = 0$ in \eqref{boltzmannTC}. This simplification leads to linear connection between the diffusion vector ${\bf d_e}$ and the temperature gradient. It leads to expression \eqref{heatcondterr}, which also follows from a approximate theory of heat conductivity and diffusion, based on the mean free path. In a more general case in the presence of the magnetic field, it leads to the simple dependence of the thermal conductivity tensor on the magnetization parameter.

 In outer layers of magnetized NSs the electric currents are substantial, and thermoelectric effects take place  \cite{blandford83}, so that the average velocity is not equal to zero any more, and heat transfer should be considered together with diffusion (see \cite{bisn2} for details). The expression for the heat flux we use is connected only with a temperature gradient, when the diffusion vector ${\bf d_e} = 0$. This approach is approximate as well, and a consistent consideration of thermoelectric processes has to be done. Nevertheless, our axisymmetric heat transfer simulations from Paper 1 are in good agreement with the ones by \cite{heat1} for core-dipolar magnetic fields.

\subsection{Magnetic field configuration}
We consider dipole and quadrupole configurations of the magnetic field, which are defined by the following formulae. For the dipole we have

\begin{equation}
{\bf B} = \frac{B_{pd}R^3_{NS}}{2}\frac{3({\bf d}\cdot {\bf r}){\bf r} - {\bf d}r^2}{r^5}
\label{dipole}
\end{equation}
where $B_{pd}$ is the value of a magnetic induction at the magnetic pole on the NS surface, ${\bf d}$ is a unit vector in the direction of the magnetic dipole, $R_{NS}$ is a NS radius.
For the quadrupole configuration, with a quadrupole momentum in the direction of $z$ axis, we have
\begin{equation}
{\bf B} = B_{pq}R^4_{NS}\bigg(\frac{r^2 - 5z^2}{2r^7}{\bf r} - \frac{{\bf e_z}z}{r^5}\bigg),
\label{quadrupole}
\end{equation}
where ${\bf e_z}$ is a unit vector along a z-axis. In the subsequent consideration we consider a combination of these two field configuration, with different values of $B_{pd}$ and $B_{pq}$, and different angles between vectors ${\bf d}$ and ${\bf e_z}$, in the envelope, and in the crust of NS.

\subsection{Equation of state and envelope model}

Density appears explicitly  in the thermal conductivity tensor \eqref{heatcond}. We have built a NS model
by solving Tolman-Oppenheimer-Volkoff equations of the hydrostatic equilibrium to get a density profile in
the crust. For the NS interior we used moderately stiff equation of state (EOS) SLy4 of \cite{EOS}, which is based on
microscopic calculations with an effective nuclear potential from \cite{nucl1}. The used EOS describes consistently  both the crust and the core. We have chosen central density $\rho_c=1\cdot 10^{15}$ g/cm$^3$. The NS mass is $M_{NS} = 1.42\,M_{\odot}$, where $M_{\odot}$ is the
 Solar mass, the inner and outer radii of the NS crust are $R_{in} = 10.59$ km at $\rho = \rho_{in} =
 2\cdot10^{14}$ g/cm$^3$ and $R_{out} = R_{NS} = 11.62$ km at $\rho = 10^{10}$ g/cm$^3$ respectively.
  We have taken into account a neutronization of the matter in the crust with effective $A$ and $Z$.
  Those values are taken from \cite{BBP1971} for the density $\rho<\rho_{drip}$ and from \cite{BPS1971}
   for $\rho > \rho_{drip}$, where $\rho_{drip}=4\cdot 10^{11}$g/cm$^3$ is the neutron drip density.

The outer envelope of the NS is a thin near-surface layer ($\sim 100$ metres in depth) of plasma, which extends from the NS crust to the radiative surface. It consists of partially degenerate electrons and non-degenerate iron nuclei. We have neglected the effects of
nonideality and quantizing magnetic fields on the EOS and assumed ideal fully ionized plasma of iron
($Z=26,\,\,A=56$) with non-degenerate non-relativistic nuclei, and degenerate relativistic electrons:
\begin{displaymath}
P = P_{n}^{(N)}+P_{d}^{(e)},
\end{displaymath}
where $P_{n}^{(N)} = n_N k_BT$ is an ion pressure, indices $"n"$, $"d"$ correspond to non-degenerate and degenerate  gases respectively. The pressure of the electrons at arbitrary degree of the degeneracy and relativism can written in terms of Fermi-Dirac integrals: 
\begin{equation}
P_{d}^{(e)} = \frac{(2m_e)^{3/2}}{3\pi^2\hbar^3\beta^{5/2}}\big(I_{3/2}(\chi,\tau) +
\frac{\tau}{2}I_{5/2}(\chi,\tau)\big)
\label{Pe}
\end{equation}
where $\beta = (k_BT)^{-1}$, $\chi = \beta\mu_{id}^{(e)}$ is the electron chemical potential,
normalized on $k_BT$, $\tau = (\beta m_ec^2)^{-1}$, and Fermi-Dirac integrals are defined as follows:
\begin{equation}
I_{\nu}(\chi,\tau) = \int_0^{\infty}\frac{u^\nu \sqrt{1 + \tau u/2}}{\exp(u-\chi) + 1}du,
\label{fermi}
\end{equation}
here $u = \beta m_ec^2(\sqrt{1+\frac{p^2c^2}{m_e^2c^4}}-1)$, and $p$ is an electron momentum. We used the analytical approximations for the Fermi-Dirac integrals from \cite{blin}. For the electron chemical potential $\chi$ we use a non-relativistic analytical approximation from \cite{Antia} with relativistic corrections adopted from \cite{PRE}.

In a thin envelope the radial temperature gradient, as well as the radial heat flux, are much larger than the azimuthal ones. 
Therefore, the heat flux approximately is assumed to be only radial through the envelope. Such approach leads to the local, one-dimensional plane-parallel model of the envelope thermal structure. Thus, the temperature distribution in an envelope region can be calculated separately from the crust. To solve the problem self-consistently, it is necessary to to find a common solution for the envelope and the crust, for a given temperature of the isothermal NS core my means, suggested by \cite{heat1}. The first step for finding this self-consistent solution is a calculation of the the relation between a surface temperature $T_s$, and a temperature at the bottom of the envelope $T_b$ with the fixed density $\rho_b$. This relation is constructed by solution of local one-dimensional heat  transfer equation, with different microphysics input. Due to anisotropic heat transfer in presence of a strong magnetic field, the $"T_s-T_b"$- relationship is a variable over the NS surface, depending on the magnetic field distribution. For non-magnetised NSs $"T_s-T_b"$ - relationships were constructed in e.g. \cite{G1983,PCY}, and in \cite{PY01,PCY07,heat5} they were calculated for magnetised NSs (in the latter paper in 2D approach). 

The thermal structure equation for the envelope reads (e.g. \cite{G1983,PY01}):
\begin{equation}
\frac{dT}{dP} = \frac{3K}{16g_s}\frac{T_s^4}{T^3},
\label{TSE}
\end{equation}
where $T_s$ is a local surface temperature, $K = K(B, \theta_B, T,\rho)$ is an effective opacity, $\theta_B$ is a magnetic field inclination angle to the normal of the surface, $g_s = GM_{NS}/(R_{NS}^2\sqrt{1 - r_g/R_{NS}})$ is the surface gravity acceleration, with approximate account of GR effects (e.g. \cite{G1983}, \cite{VR1988}), $G$ is the gravitational constant, $r_g$ is the NS gravitational radius.

The heat flux is determined by a sum of two processes: radiative and electron heat transfer. For the radiative opacity we have taken into account free-free and bound-free transitions as well as an electron Thompson scattering for both non-degenerate and degenerate electrons. More detailed discussion can be found in Appendix A. The electron opacity can be derived from an analogy with a radiative heat transfer:
\begin{equation}
K_e = \frac{16\sigma T^3}{3\kappa_e\rho}.
\label{opacity_e}
\end{equation}
Here $\kappa_e = \kappa_{e\parallel}\cos^2\theta_B + \kappa_{e\perp}\sin^2\theta_B$ is a local effective value of the thermal conductivity coefficient \eqref{heatfl} for degenerate electrons. 

Equation \eqref{TSE} is solved as a Cauchy problem for the given values of surface temperature $T_s$ and surface pressure $P_s$. The latter is calculated from Eddington approximation $P_s\approx \frac{2g_s}{3K(B, \theta_B, T_s,\rho_s)}$, see \cite{PCY}. 
We used the tabulated $T_s-T_b$-relationship for $\rho_b = 10^{10}$ g/cm$^3$ to implement it in a radiative outer boundary condition for the heat transfer equation in the crust, which was presented and discussed in detail in Paper 1.

\section{A boundary-value problem}

 After a fast stage of
neutrino cooling, the core of the NS cools down slowly, and temperature distribution may be considered as a
stationary one. So that, a thermal evolution of the NS could be considered as a sequence of cooling models with a stationary temperature distribution over the NS.
 The temperature is supposed to be constant through the core $T_{\rm core}$, because of a large value of the heat conductivity, and to be equal to the value on the inner radius of the crust.

In a thin envelope the radial temperature gradient, as well as the radial heat flux, are much larger than the azimuthal ones, which are not considered subsequently any more. In a thin low-mass envelope the local heat flux $F_s$,  from the NS unit surface, is supposed to be constant along the radius, varying only over the surface. The flux is also supposed to have a black-body spectrum, and follows the Stephan-Boltzmann law $F_s(\theta, \phi) = \sigma T_s^4$, with $T_s(\theta, \phi)$ being the local surface NS temperature.

For a given core temperature $T_{\rm core}$, the way of construction of thermal NS model, and its surface temperature $T_s(\theta, \phi)$ may be summarized in the following way.

 1. Take a trial value of the surface temperature $T_s^{(1)}$, which determines uniquely the surface parameters, and local heat flux $F_s^{(1)}$. Using these parameters as boundary conditions for solution of the heat conduction equation in the envelope, we obtain the value of the temperature at the bottom of the envelope $T_b^{(1)}(\theta,\phi)$, and find $T_s$ - $T_b$ relationship.

2. The value is given for the core temperature $T_{\rm core}$ which is equal to the temperature at the inner crust with $r = R_{in}$, and $r = R_{out}$, as inner and outer radii of the crust. The local flux distribution $F_s^{(1)}(\theta, \phi)$ at the outer crust boundary is taken from the envelope structure. The conditions at the inner and outer boundaries are

\begin{equation}
T_{in} = T_{\rm core}, \quad \kappa ({\bf B}, \rho, T)\nabla_r{T}|_{out} + F_s = 0,
\label{bcore}
\end{equation}
In the spherical layer $R_{in}\le r \le R_{out}$ we solve the boundary-value
problem for the heat transfer equation

\begin{equation}
\nabla\cdot \kappa ({\bf B}, \rho, T)\cdot\nabla{T} = 0
\label{heattr}
\end{equation}
with the boundary conditions \eqref{bcore}.
We calculate 3D model of the magnetized crust, and obtain distribution of the temperature on the outer crust boundary $T^{(1)}_{cr}(\theta, \phi)$.

3. In the self-consistent model the temperature at the inner boundary of the envelope should coincide with the temperature at the outer boundary of the crust,  so two distributions should coincide

\begin{equation}
 T_b(\theta, \phi)= T_{cr}(\theta, \phi).
 \label{bcore0}
\end{equation}
The iterations by Newton method are performed until the equality  $T_b^{(i)}(\theta, \phi)=
T_{cr}^{(i)}(\theta, \phi)$ will be fulfilled with a necessary precision. This procedure should be
performed for each magnetic field distribution under investigation.

A heat transfer problem in the crust of a magnetized NS was solved numerically with our extension of
the Basic operators method (\cite{ard1}). 3D mesh analogues of main differential operators on an unstructured
tetrahedral mesh were derived by \cite{kondmois}, numerical method for obtaining a self-consistent solution
of the heat transfer equation in the crust is developed in Paper I, the numerical implementation of the method is briefly discussed in Appendix B of this paper.

\section{Results}
\subsection{Temperature distributions}

In the Paper 1 we have calculated a temperature in the NS crust and on its surface for pure-dipolar,
pure-quadrupolar magnetic fields and their coaxial superpositions.  In this work we have obtained the
temperature distributions for non-coaxial superpositions of core-dipolar and quadrupolar fields and have
built thermal black-body light curves, which correspond to the obtained temperature distributions.
Because there is no physical constraint, which prohibit the rotation of one field multipole from another
one, inclusion of a quadrupolar component in addition to the dipolar one leads us to consideration of 2
more physical parameters, which affect the spatial temperature distribution: an angle between the dipolar
and quadrupolar components $\Theta_b$ and a relation between the polar inductions of the components $\beta =
B_{pq}/B_{pd}$, which determines the $"$strength$"$ of 3D effects. The first (and obvious) conclusion is,
that if $\beta \ll 1$, then the temperature approaches to a pure-dipolar configuration, and when $\beta
\gg 1$, a pure-quadrupolar one is observable.

\begin{figure}
\centering
	{\includegraphics[width=7.5cm,height=6.5cm]{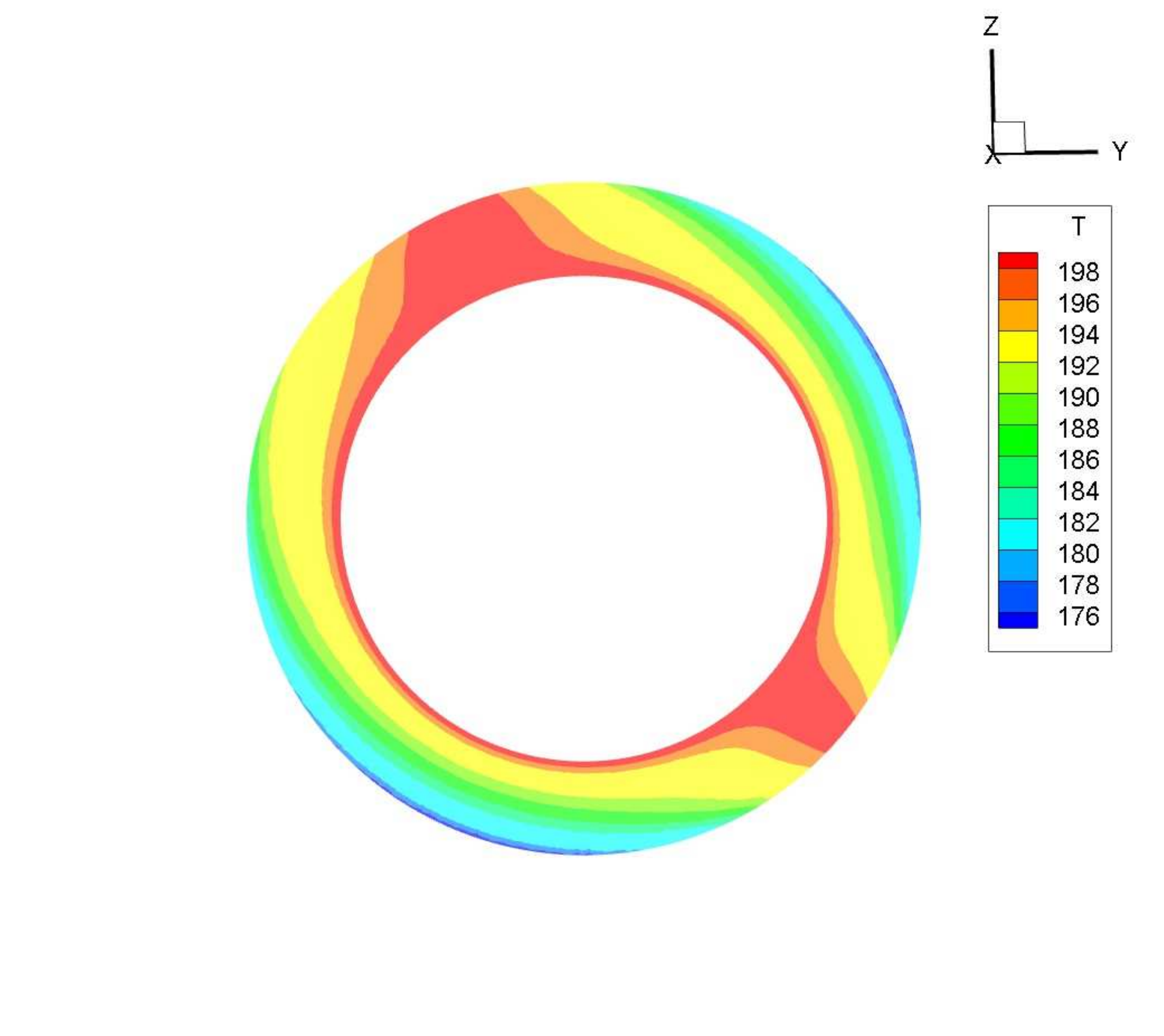}
	\includegraphics[width=7.5cm,height=6.5cm]{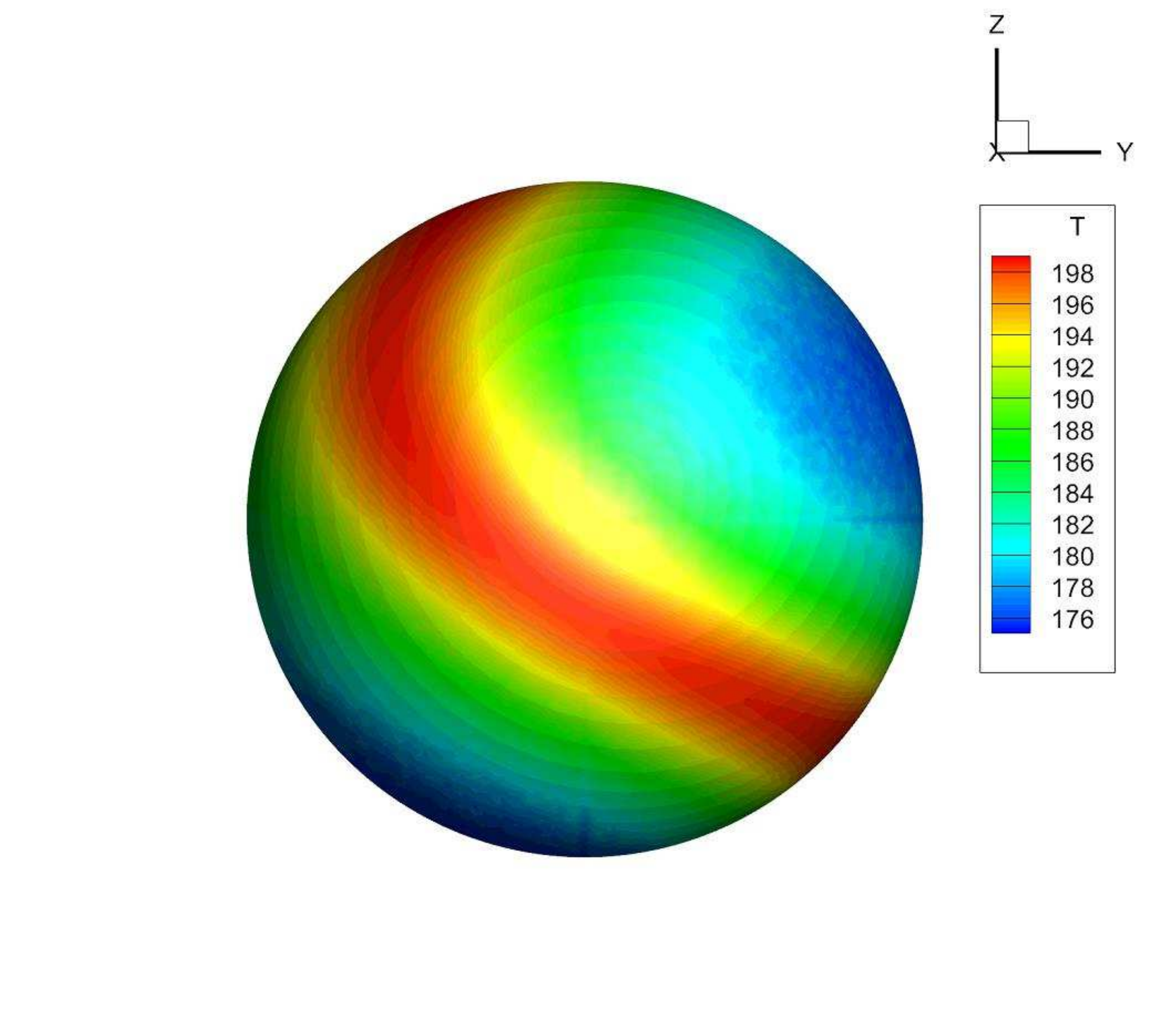}}\\
	\caption{  Temperature distribution (in units of $10^6K$) in the NS crust for quadrupolar and dipolar
 magnetic fields with polar inductions $B_{pq} = 5\cdot10^{12}\,$G and $B_{pd} = 1\cdot10^{13}$G
correspondingly ($\beta = 0.5$). Magnetic axes are rotated from each other to the angle $\Theta_b  = \pi/4$,
hereinafter a quadrupolar axis if fixed along Z-axis, and dipolar component is rotated on plots. The core
temperature is $T_{core} = 2\cdot10^8K$.  Upper picture - cross-section in Z-Y plane (the thickness of the crust is 4 times stretched for better visualization), lower one - the NS
crust surface.}
	\label{fig:Tcrust}
\end{figure}

On the Fig. \ref{fig:Tcrust} temperature distribution in the NS crust is shown for magnetic the dipole and
quadrupole, which are rotated on an angle $\Theta_b  = \pi/4$ from each other, and the quadrupolar strength
at the quadrupole magnetic pole is a half from the dipolar one, $\beta = 0.5$. The crustal temperature
distribution is inverted in comparison to the surface one, i.e. the crust temperature is smaller in regions, where the magnetic field is at least radial, and larger in the regions with an almost tangential field.
The cause is as follows. The heat flux is suppressed most crucially in the envelope, where the magnetization
parameter $\omega\tau \gg 1$. The suppressed heat flux from the envelope in the NS regions with the tangential
field (equatorial regions) causes decrease of a temperature gradient in the crust, so that a variation of the
crust temperature on the magnetic poles, where the field lines are radial, is higher, than on the equator.
A temperature variation in the crust is less than 20\% of its value.

\begin{figure}
	\centering
	{\includegraphics[width=7.5cm,height=6.5cm]{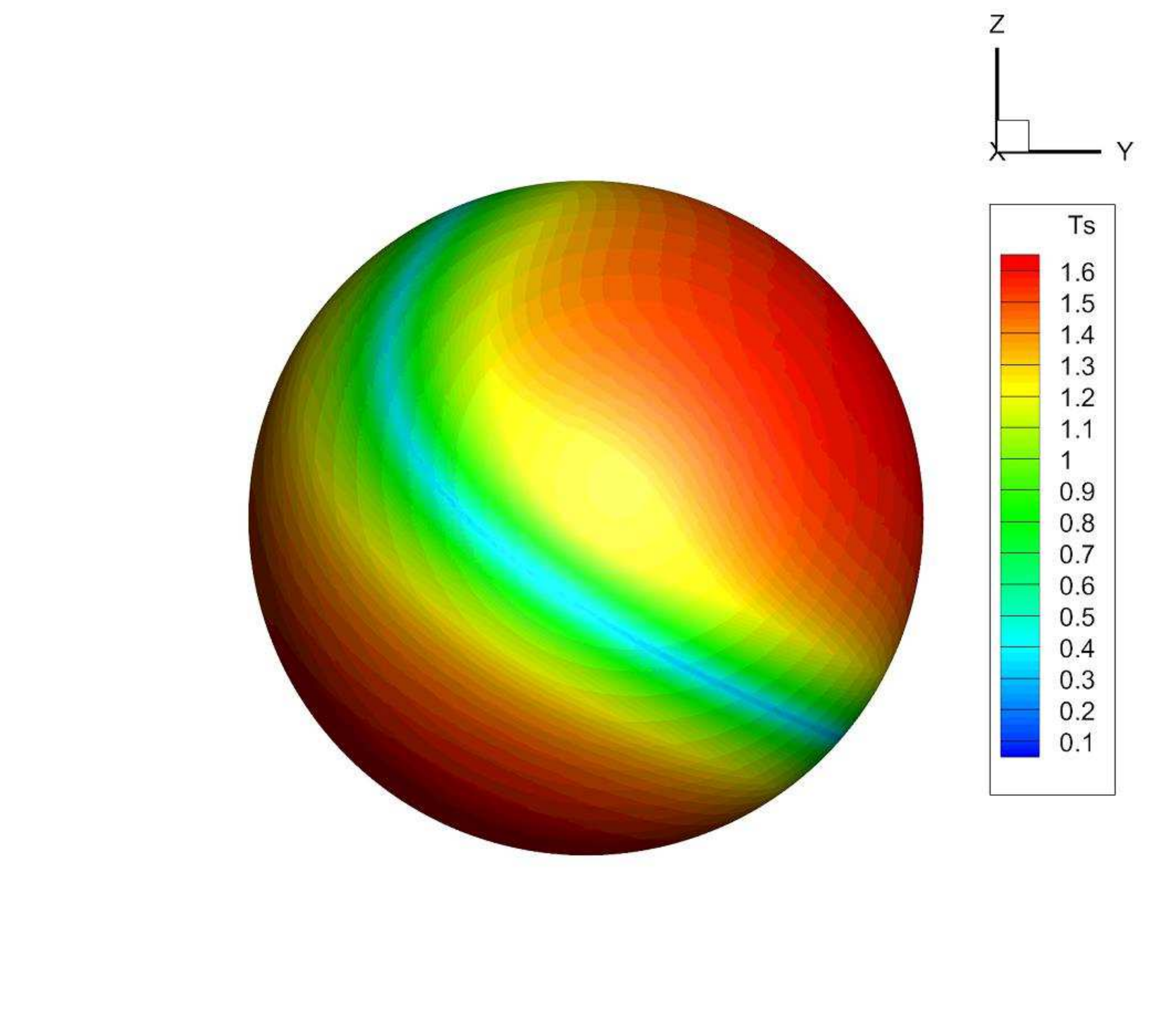}
	 \includegraphics[width=7.5cm,height=6.5cm]{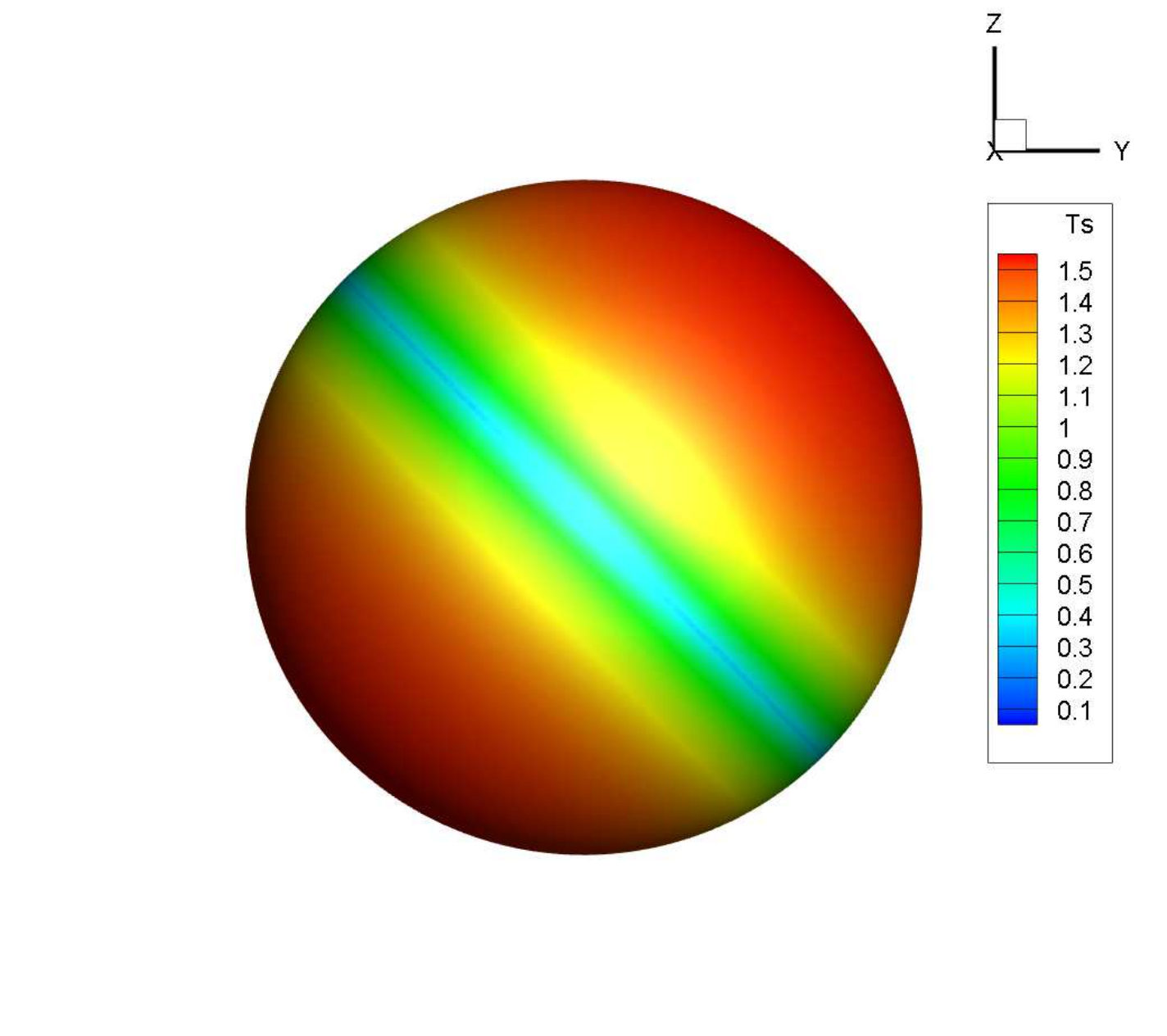}}\\
	\caption{Surface temperature distribution (in units of $10^6K$) for the same parameters as on Fig.\ref{fig:Tcrust} (upper panel) and for the NS without the quadrupolar field (lower panel).}
	\label{fig:Tsurf1}
\end{figure}

The surface temperature distribution, which corresponds to the crustal temperature from the
Fig. \ref{fig:Tcrust}, is presented on the Fig. \ref{fig:Tsurf1} (upper panel) together with the surface
temperature of the NS without the quadrupole field (lower panel). A minimal temperature is approximately
$3\cdot10^5K$, and a maximum one is near $1.6\cdot10^6K$. In a pure-dipole case the surface temperature
distribution is represented by two hot polar caps and a cold ring-shaped $"$belt$"$. A $"$switching on$"$
of the quadrupolar field effects on the heat transfer as follows. If the parameters $\beta<\sim1$ and
$\Theta_b \neq 0$, the belt shape becomes irregular, and also the belt is broaden from the one side
in comparison to the pure-dipolar configuration. Hot polar caps aren't located in antipodal positions in
such case,and they have different sizes as well, resembling RX J0720.4–3125 (\cite{obs3}). The presence
of the quadrupolar field decreases slightly an effective temperature of the NS. Thus, the cold region is
larger, than in a pure-dipolar case.

\begin{figure}
	\centering
	{\includegraphics[width=7.5cm,height=6.5cm]{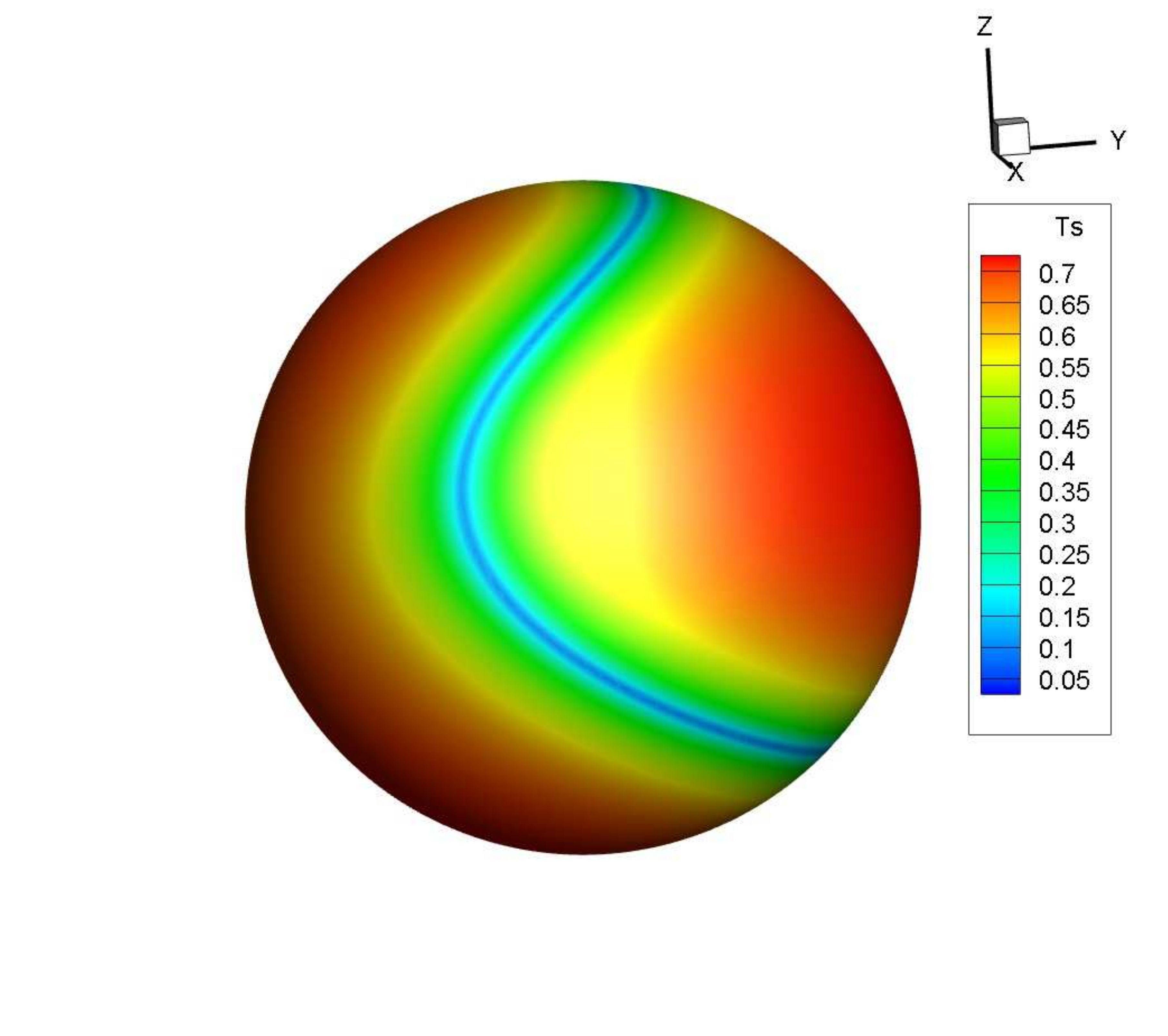}
	 \includegraphics[width=7.5cm,height=6.5cm]{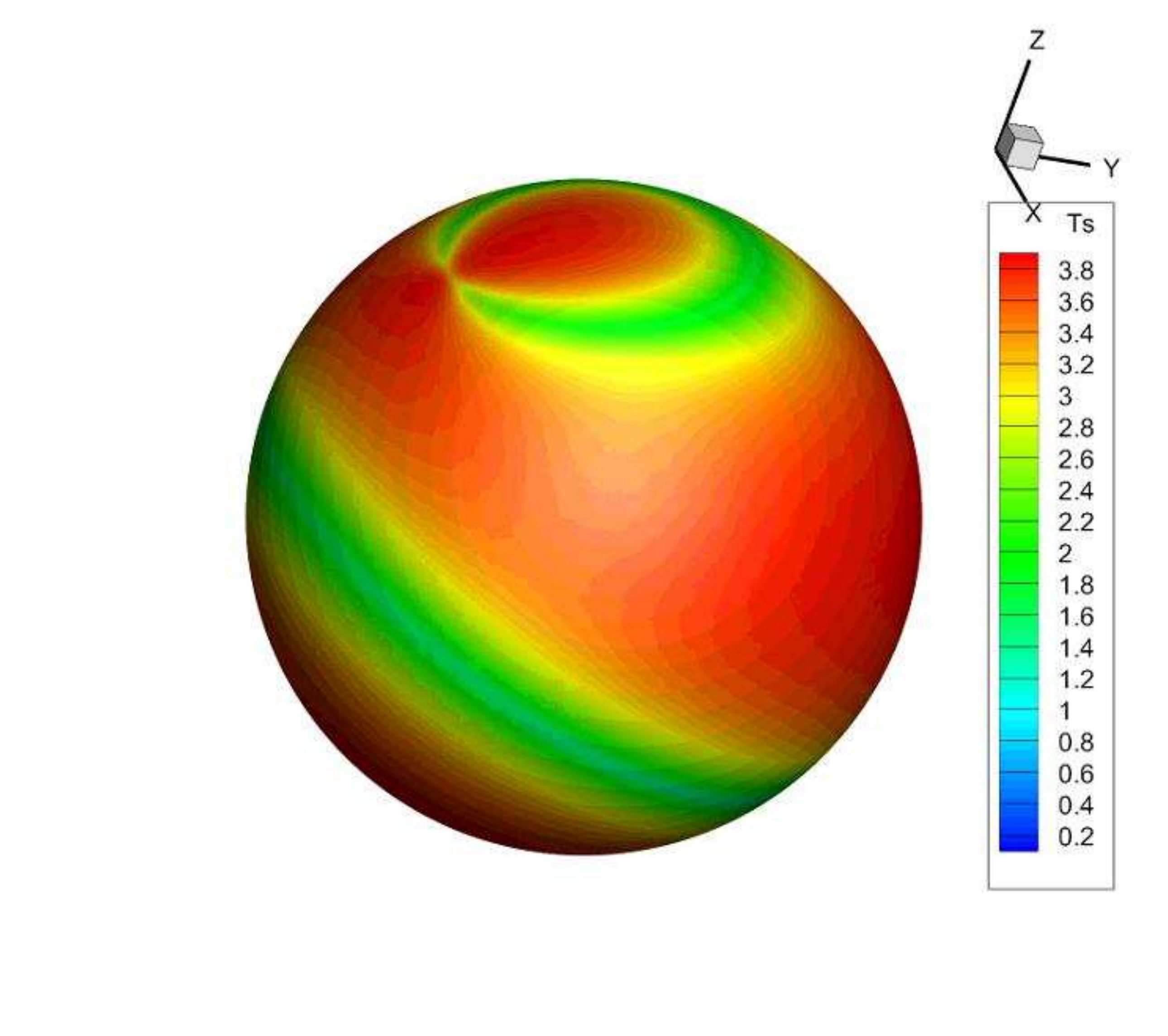}}
	\caption{Surface temperature distributions (in units of $10^6K$) with parameters $\beta=0.75$, $T_{core}
 = 5\cdot10^7K$ and $\Theta_b  = \pi/3$ (upper panel) and $\beta=1$, $T_{core} = 5\cdot10^7K$ and $\Theta_b
  = \pi/6$ (lower panel). Dipolar component $B_{pd} = 1\cdot10^{13}$ in both cases.}
	\label{fig:Tsurf23}
\end{figure}
On the Fig.~\ref{fig:Tsurf23} the temperature distributions are shown for the different $\beta$ and $\Theta_b$
 as well as for different core temperatures $T_{core}$, which correspond to various NS ages.
With increasing of the quadrupole field, the second belt appears, when $\beta \sim 0.9$ for the moderate angles
 $\Theta_b \sim \pi/6$, and on practice, when $\beta \gtrsim 1.5$, the temperature approaches to the
 pure-quadrupolar one. For the angles $\Theta_b \lesssim \pi/2$ the shape of the belt takes away more and more
from a circular one with increasing of $\beta$, getting a shape of a $"$jaw$"$, and for
$\beta \gtrsim 1.5$(1.5
 for $\Theta_b =\pi/3$) it transforms into the shape of two belts. Also we note here, that our numerical
  solution is even with respect to a secant plane, that passes through the dipolar and quadrupolar axes.

 According to 2D cooling calculations by \cite{heat4}, $T_{core} = 2\cdot 10^8K$ and $5\cdot10^7K$ correspond to NS ages $\sim 500$ years and $\sim 10^4-10^5$ years, while $T_{core} = 1\cdot10^9K$ corresponds to the NS age $\sim 1$ year, due to NS cooling by neutrinos emitted from the NS core, which don't disturb the radiation flux. Neutrino emitted from the crust itself could change its temperature distribution, and influence the thermal flux. Our calculations with the highest core temperature ($T_{core} = 1\cdot10^9K$) should be considered as model example, because we have not included neutrino losses in the crust, although they are not-negligible. Nevertheless, the surface temperature distribution will not be distorted significantly. The neutrino losses in the crust at such temperatures may cause some redistribution of the crustal temperature in accordance with inclusion of different cooling processes on $\rho-T$ plane, but the part of the neutrino flux emitted from the crust is much smaller than the NS radiation flux, so that the surface temperature is not affected significantly by neutrinos as well as the shape of its thermal light curve. 
 
During a cooling of the NS, the surface temperature anisotropy is amplified. Thus, when the core temperature
$T_{core} = 1\cdot10^9K$, a ratio between the hottest and coldest temperatures is $T_h/T_c\sim 2.5$, and
when the core temperature cools to $5\cdot10^7K$, it equals to $T_h/T_c\sim 7$. The magnetization parameter $\omega\tau$ is weakly temperature-independent, and thus we have obtained in our assumptions, that the
temperature distributions have the same topology during a NS thermal evolution, if the magnetic dissipation
effects are not included.

\begin{figure}
	\centering
	\includegraphics[width=9.0cm,height=4.5cm]{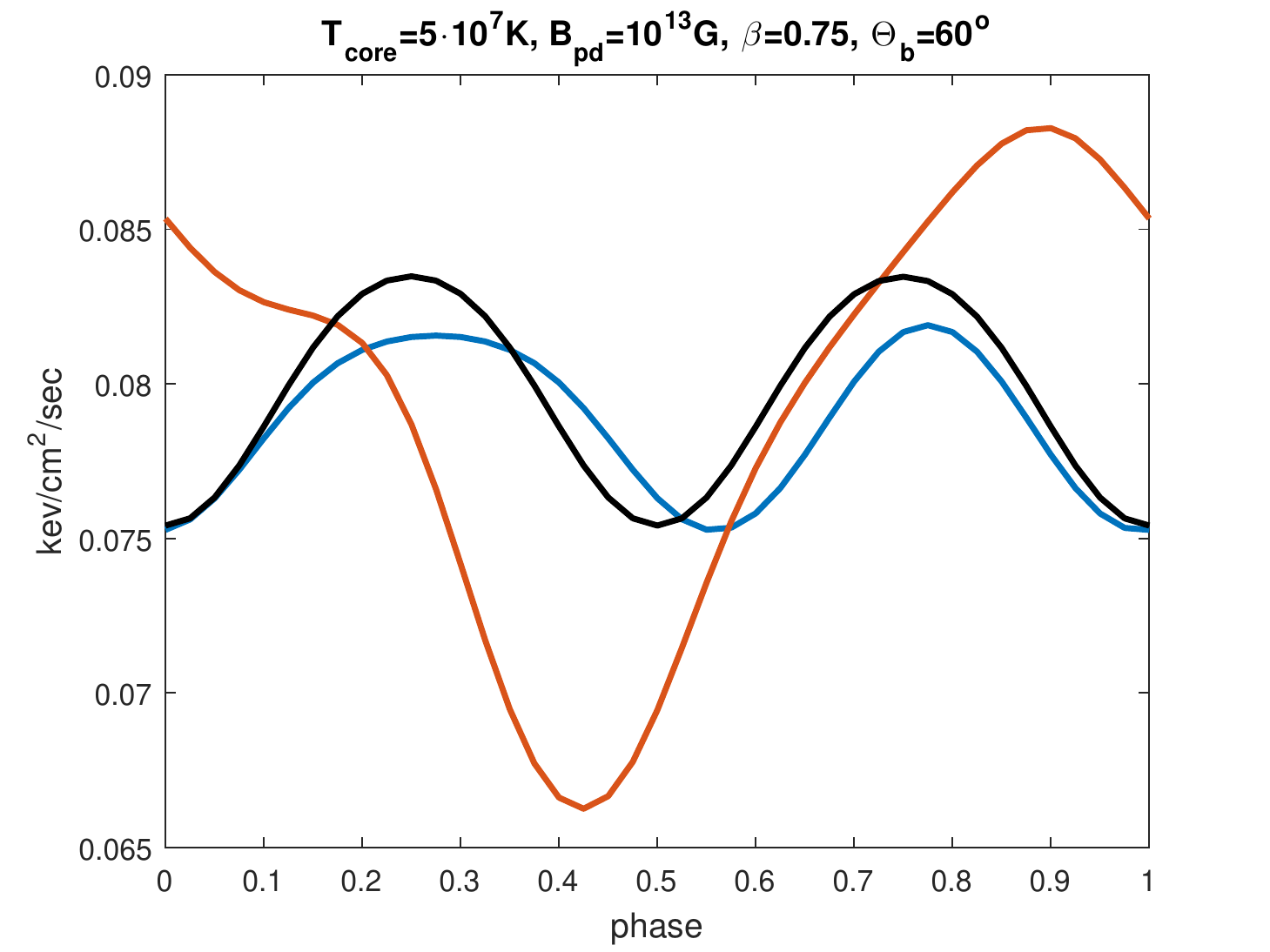}
	\includegraphics[width=9.0cm,height=4.5cm]{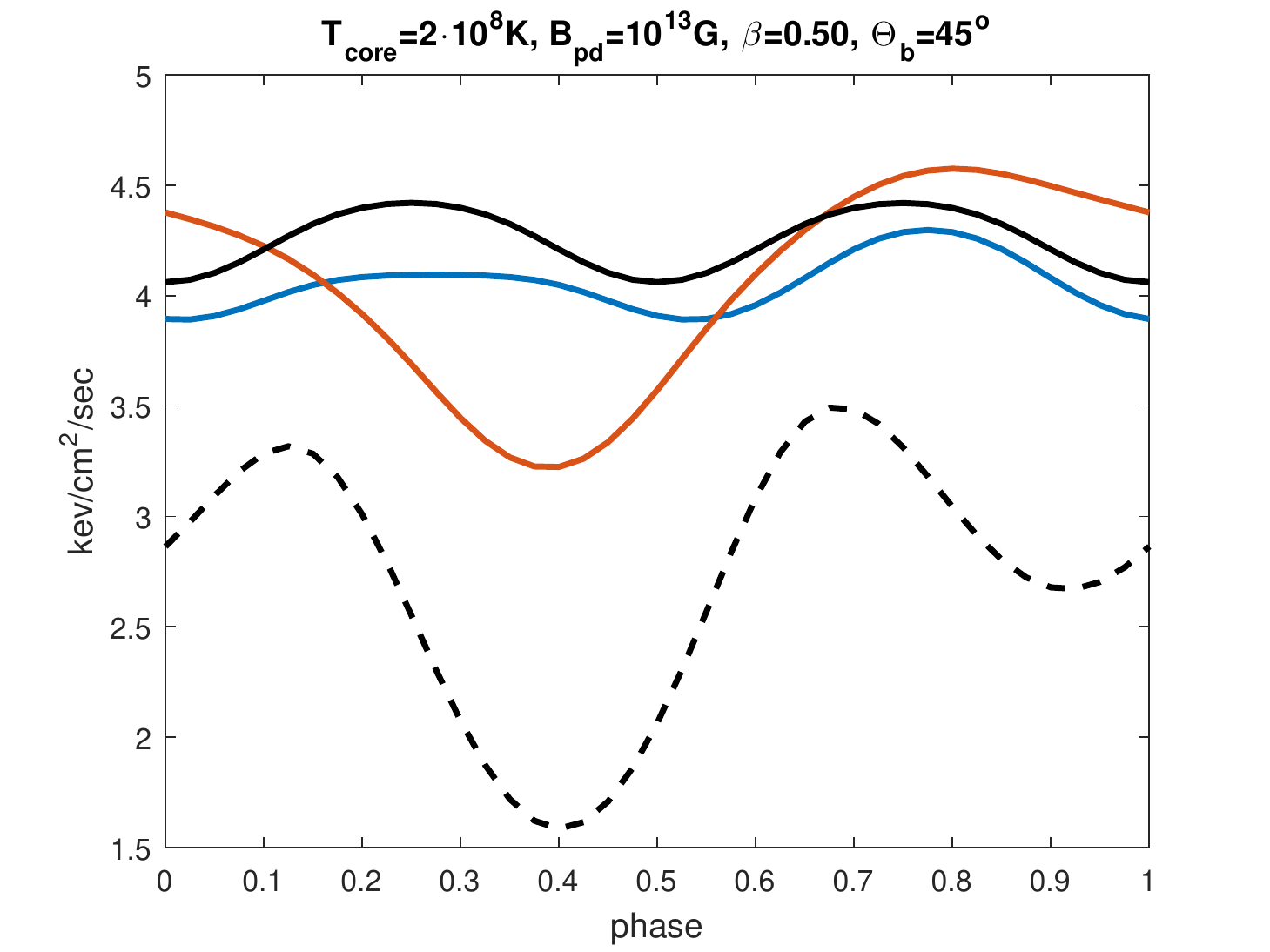}
	\includegraphics[width=9.0cm,height=4.5cm]{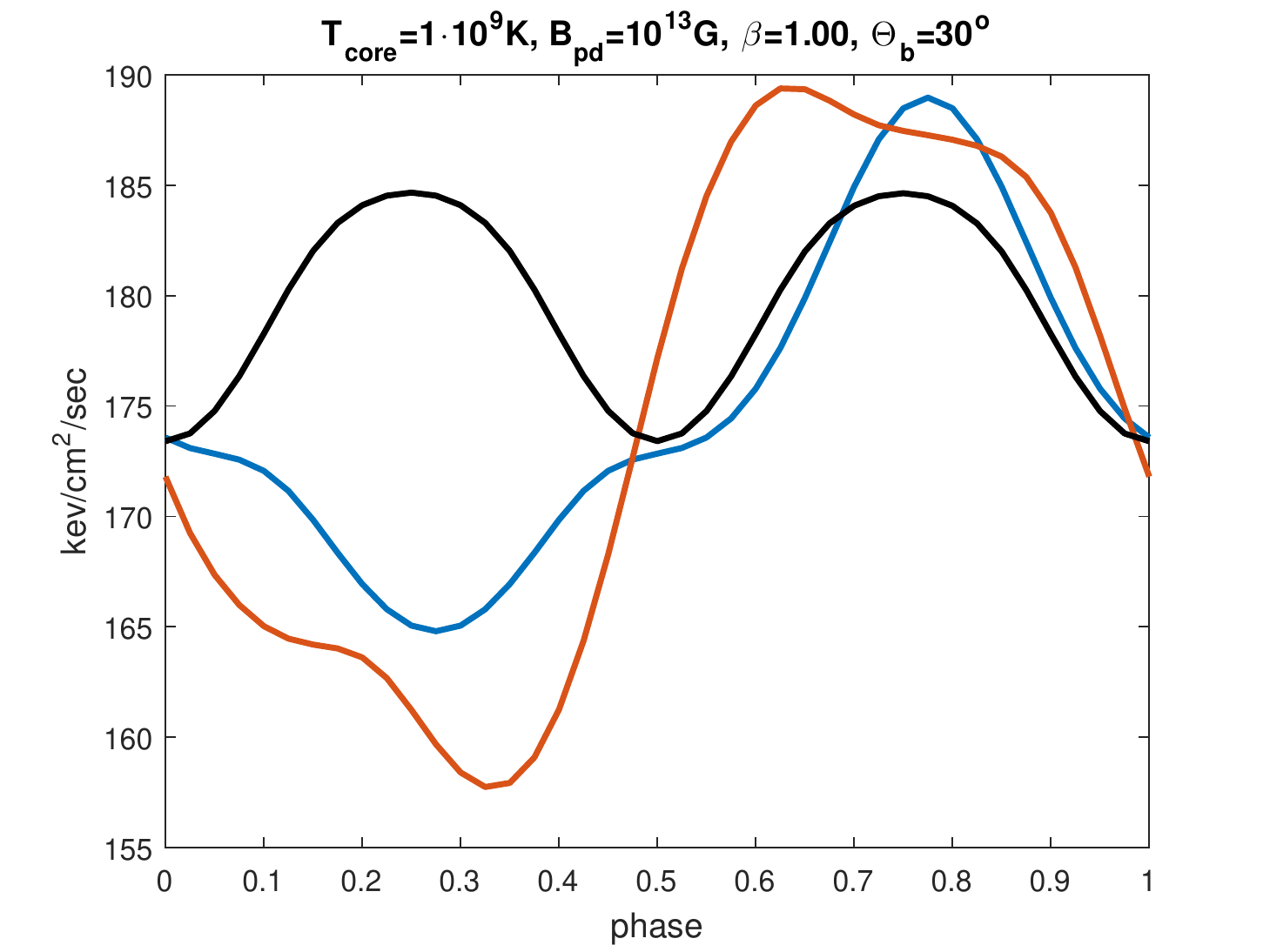}
	\caption{Light curves for the different $T_{core}, \beta$ and $\Theta_b$ (ones from Figs.\ref{fig:Tsurf1},\ref{fig:Tsurf23}) for an orthogonal rotator ($\alpha_d = \zeta = \pi/2$). Blue lines correspond to the case, when rotational, dipolar and quadrupolar axes are in the same plane, red ones correspond to the case, when $\alpha_q = \alpha_d$, so both magnetic axes are visible for the observer. Black lines correspond to the light curves from the NS with a purely dipolar field. A dashed black line is the light curve from the Case 2, but without taking into account light bending effects.}
	\label{fig:lc2p}
\end{figure}

\subsection{Observational manifestations}

A modelling of thermal light curves from the rotating magnetized NSs (particularly from XDINSs and X-ray
pulsars) is a pretty studied topic. A thermal emission from compact object was considered by a long list
of authors (e.g. by \cite{GH1983} without general relativity, and by \cite{pfc1983, p1995, ps1996,
zaneturolla,turolla13} with taking into account light bending effects). We have examined observational
manifestations from the obtained surface temperature distributions using a simple composite black-body model.
The thermal energy flux is defined in a Newtonian space-time as follows:

\begin{equation}
\begin{gathered}
dF = \frac{2\pi}{c^2h^3}\frac{R_{NS}^2}{D^2}\bigg[\int\displaylimits_{\cos\theta>0}dS\cos\theta I_E(E,T(\theta, \varphi))\bigg]\times \\
\times E^3\exp^{-N_H\sigma(E)} A(E) dE dt,
\end{gathered}
\label{fluxwogr}
\end{equation}
where $I_E = \big(\exp\big(\frac{E}{k_BT}\big)-1\big)^{-1}$ is Planck distribution function, $E$ is a photon
energy, $\theta$ and $\varphi$ are azimuth and polar angles correspondingly, $D$ is a distance from the NS to
an observer and $A(E)$ is an effective detector area. We do not consider a detector response in this paper
and assume, that the detector has a unit area. An interstellar absorption is taken into account by the
term $\exp{(-N_H\sigma(E))}$, where $N_H$ is a hydrogen column density between the NS and observer, and
$\sigma(E)$ is an effective absorption cross-section (\cite{absorb}). Throughout this part of our work we
consider $N_H = 10^{20} cm^{-2}$. An expression in square brackets in \eqref{fluxwogr} has to be integrated
over the surface of a visible hemisphere, and after that $dF$ is integrated over the energies of the photons
to obtain a phase-dependent light curve.

General relativistic effects are sufficient for the compact objects. A rigorous relativistic theory of a light
propagation near the compact object was developed by \cite{pfc1983}. In real conditions of the NS the effects
of the general relativity are pronounced mostly by a redshift of the photon energy and a deviation of the
photon trajectory from the straight line. The latter effect manifests itself as follows. The ray, which
leaves the surface with an angle $\theta'$ to the normal to the NS surface, will be bend, and at the infinity
this angle will be $\theta>\theta'$ for the observer. Thus, more than a hemisphere is observable, and an
effective visible NS radius is more, than the exact one. A simple, but good approximate formula for the
relation between $\theta$ and $\theta'$ was proposed by \cite{bel2002}, and we use it in further calculations:

\begin{equation}
1 - \cos\theta' = (1 - \cos\theta)(1- x_g),
\label{belob}
\end{equation}
where $x_g = \frac{2GM_{NS}}{c^2R_{NS}}$. With the inclusion of the described effects in the considered
model, the energy flux is written as follows:

\begin{equation}
\begin{gathered}
dF = \frac{2\pi}{c^2h^3}\frac{R_{NS\infty}^{2}}{D^2}\bigg[\int\displaylimits_{\cos\theta'>0}dS\cos\theta' I_E(E_{\infty}(1-x_g)^{-1/2},T)\bigg]\times \\ \times E_{\infty}^3\exp^{-N_H\sigma(E_{\infty})} A(E_{\infty}) dE_{\infty} dt_{\infty}.
\end{gathered}
\label{fluxgr}
\end{equation}
In the formula above $E_{\infty} = E\sqrt{1-x_g}$ is a redshifted energy, an effective NS radius is equal to
$R_{NS\infty} = R_{NS}/\sqrt{1-x_g}$,
 and term $dt_{\infty} = dt/\sqrt{1-x_g}$ corresponds to a time dilation near the
NS. The value of $\theta'$ is obtained from \eqref{belob}, and the integration of the expression in the
square brackets should be done over the visible part of the surface. We consider an energy band of
the XMM-Newton EPIC-pn detector, so that $E_{min} = 0.15$keV, and $E_{max} = 1.5$keV.

During observation of the thermal emission from the rotating magnetized NS, pulsations of the visible flux arise.
To measure their strength, let us introduce a so-called pulsed fraction ($PF$):
\begin{equation}
PF = \frac{F_{max} - F_{min}}{F_{max} + F_{min}},
\label{mpf}
\end{equation}
where $F_{max}$ and $F_{min}$ are the values of the maximal and minimum fluxes of energy (the fluxes of
photon counts may be considered for the X-ray sources as well).

\begin{table}
	\caption{$PF$ (\%) which correspond the two-peaked light curves from the Fig.\ref{fig:lc2p}. }
	\label{tab:table1}
	\begin{center}	
		\begin{tabular}{ | c | c | c |c| }	
			\hline
			& $\beta=0.75$ & $\beta=0.50$& $\beta=1.00$ \\
			& $\Theta_b=60^o$ & $\Theta_b=45^o$& $\Theta_b=30^o$ \\ \hline
			dipole & 5.1 & 4.2 & 3.1 \\
			Case 1 & 4.2& 4.9 & 6.8 \\
			Case 2 & 14.2 & 17.2 & 9.1 \\
			
			\hline
		\end{tabular}
	\end{center}
\end{table}	
In the absence of the quadrupolar component the pulse profile is symmetric and sinusoidal, and light curve
can be either two-peaked (both magnetic poles are visible) or one-peaked (one precessing pole is visible).
Also we introduce here two angles, which characterize a light curve: an angle between the rotational an the
dipolar axes $\alpha_d$, and an angle between the rotational axis and a line of sight of the observer $\zeta$.
It was noticed by \cite{GH1983}, that for pure-dipolar magnetic field configurations, when $\alpha_d +
\zeta\leq\pi/2$, then the light curve is one-peaked, and else, it is two-peaked. General relativistic
effects make this conclusion more strict (\cite{p1995}). Inclusion of the quadrupolar component adds one
more degree of freedom in a space of positions for the axes, which characterize the light curve, so it makes
its analysis much more complicated. Thus, we consider only two limits: the first case (Case 1) is when all
three axes - rotational, dipolar and quadrupolar - are in the same plane, and the second one (Case 2) is
when both dipolar and quadrupolar axes are on the line of sight of the observer in some moments of time
during rotational period, so that $\alpha_d = \alpha_q$, where $\alpha_q$ is an angle between the rotational
and quadrupolar axes (e.g. when magnetic axes are in the equatorial plane with respect to the rotational one).
Also, we restrict ourselves with a constraint $\alpha_d = \zeta$, unless otherwise specified.

On the Fig.\ref{fig:lc2p} the light curves for the orthogonal rotator ($\alpha_d = \zeta = \pi/2$) are
presented for the different temperature distributions from the previous subsection for the both limits for
the positions of the axes and for the pure-dipolar magnetic field configurations (black lines). When all
three axes are in the same plane (blue lines, Case 1), the main indicator of an existence of the quadrupolar
field is as follows. The light curves changes slightly from the dipolar ones in the absence of the second
cold belt in the temperature distribution (the first two pictures). One peak is tighten, and the second one
is broaden in comparison to the pure-dipolar case. $PF$ decreases slightly, e.g. it is $5.1\%$ for the
pure-dipole field and $PF = 4.2\%$ for the first picture on the Fig.\ref{fig:lc2p}, then $T_{core} =
5\cdot10^7K$. Table \ref{tab:table1} gives information about the maximum pulsed fractions for the different
light curves. Pulse profiles are symmetric in the Case 1, because of the parity of the solution with respect
to the secant plane, which is built on the magnetic axes, so that the light curve has a mirror symmetry at
a half period. The third picture from the Fig.\ref{fig:lc2p} corresponds to the temperature distribution,
where the second cold belt appeared, so that an additional lay-down is observable instead of the second peak.

More interesting situation is provided for the Case 2 (red lines): the symmetry of the pulse profiles is
broken, and light curves can take various shapes. Moreover, the pulsations are amplified sufficiently, from
4\% up to 17\% on the upper panel of Fig.\ref{fig:lc2p}. Perhaps, a red line on this picture from
Fig.\ref{fig:lc2p} may describe qualitatively the light curve from RX J0420.0-5022 (\cite{Haberl04}).
Its $PF=14.2$\%, and a pulse shape is close to the one observed by the XMM-Newton. Also, it should be
mentioned, that all the synthesized light curves have become one-peaked due to the effects of general
relativity. A dashed line on the middle panel of Fig.\ref{fig:lc2p} correspond to the light curve in a
flat space-time, and it is two-peaked. The light bending effects have "blurred" the pulse profile, making
two peaks merge into one.

\begin{figure}
	\centering
	\includegraphics[width=9.0cm,height=4.5cm]{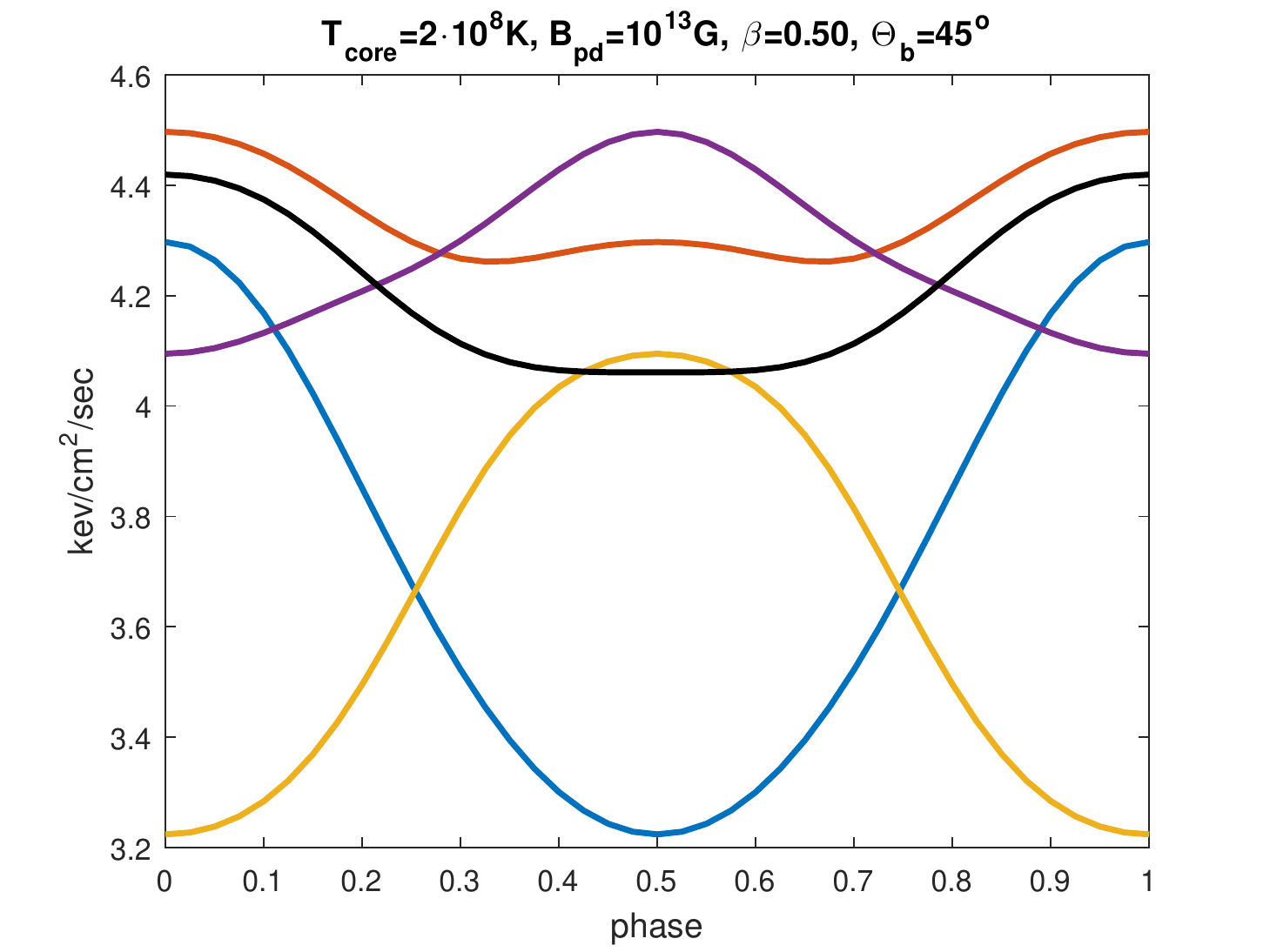}
	\caption{One-peaked light curves for the case, when all axes are in the same plane for the different
positions of the magnetic component to each other. Physical parameters are: $T_{core} = 2\cdot10^8K$,
$B_{pd}=10^{13}G, \beta = 0.50$ and $\Theta_b=45^o$ (Fig.\ref{fig:Tsurf1}). Black lines correspond to
the pure-dipole. Coloured lines correspond to the different positions of the quadrupolar component: blue
and yellow lines correspond to the smaller polar cap, while red and violet ones show the light curves for
the larger polar cap. }
	\label{fig:lc1pni}
\end{figure}

We have considered the effect of the non-coaxial quadrupolar field on the one-peaked light curves, i.e. on
the curves, on which only one polar cap is visible for the pure-dipolar magnetic field. We still consider
two limits for the positions of the quadrupolar axis in the one-peaked case as well as for two peaked light
curves. On the Fig.\ref{fig:lc1pni} the light curves for the Case 1 are presented for different positions of
the quadrupolar field. In comparison to the light curve in the pure-dipolar case (black line), a presence
of the quadrupole requires to consider more cases of position of the observer and the quadrupole. The
quadrupolar field makes one hot polar cap $"$smaller$"$ (those one, where dipolar field lines directed
$from$ the NS surface $to$ the core), while the second cap remains at least the same or becomes larger due
to a shift of a belt (\cite{ps1996}; Paper 1) and its curvature, so that the caps are distinguishable, one
from each other. If $\Theta_b=0$, then only two types of the light curves describe emission from poles, if $\alpha_d =
\zeta$, and when the angle $\Theta_b$ between components is not equal to zero, the position of the quadrupole with
respect to the dipole and the observer leads to four different types of the light curves. All the synthesized light curves are
symmetric, and $PF$ can be amplified up to $14.7\%$ for the smaller cap in comparison to the $PF = 4.2\%$
for the pure dipole, and the larger cap is at least indistinguishable for the observer from the pure-dipolar
one by its thermal emission. Moreover, $PF$ can be amplified by different physical processes (e.g. by
inclusion of an absorption line in the spectrum, see review by \cite{turolla09} about XDINSs as well as a paper by \cite{magn8}, where the recently discovered and fitted thermal component of PSR J0726-2612 shows strong pulsations), so that an absence of quadrupolar
features on the curves may make their analysis more difficult.

\begin{figure}
	\centering
	\includegraphics[width=9.0cm,height=4.5cm]{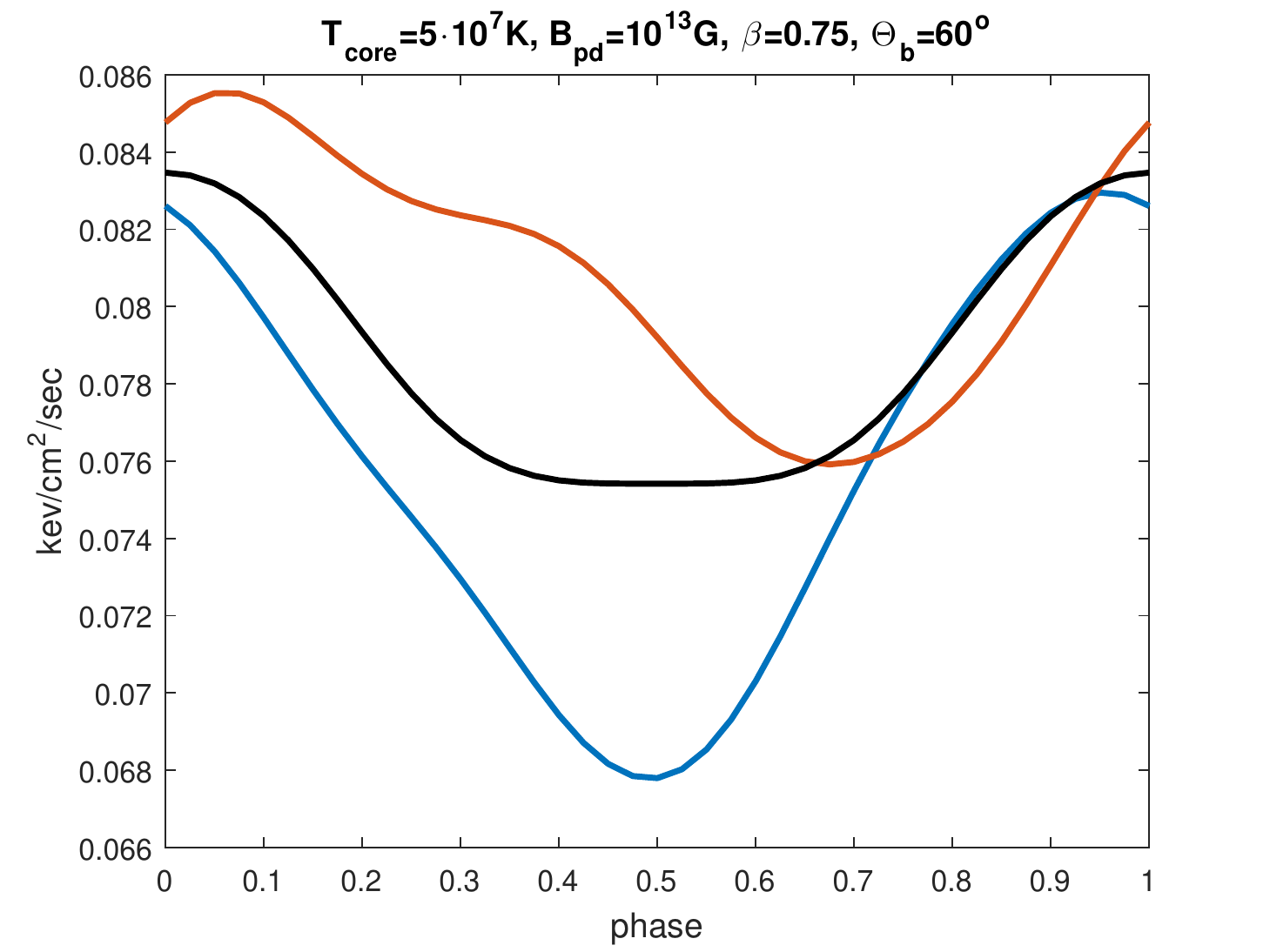}
	\includegraphics[width=9.0cm,height=4.5cm]{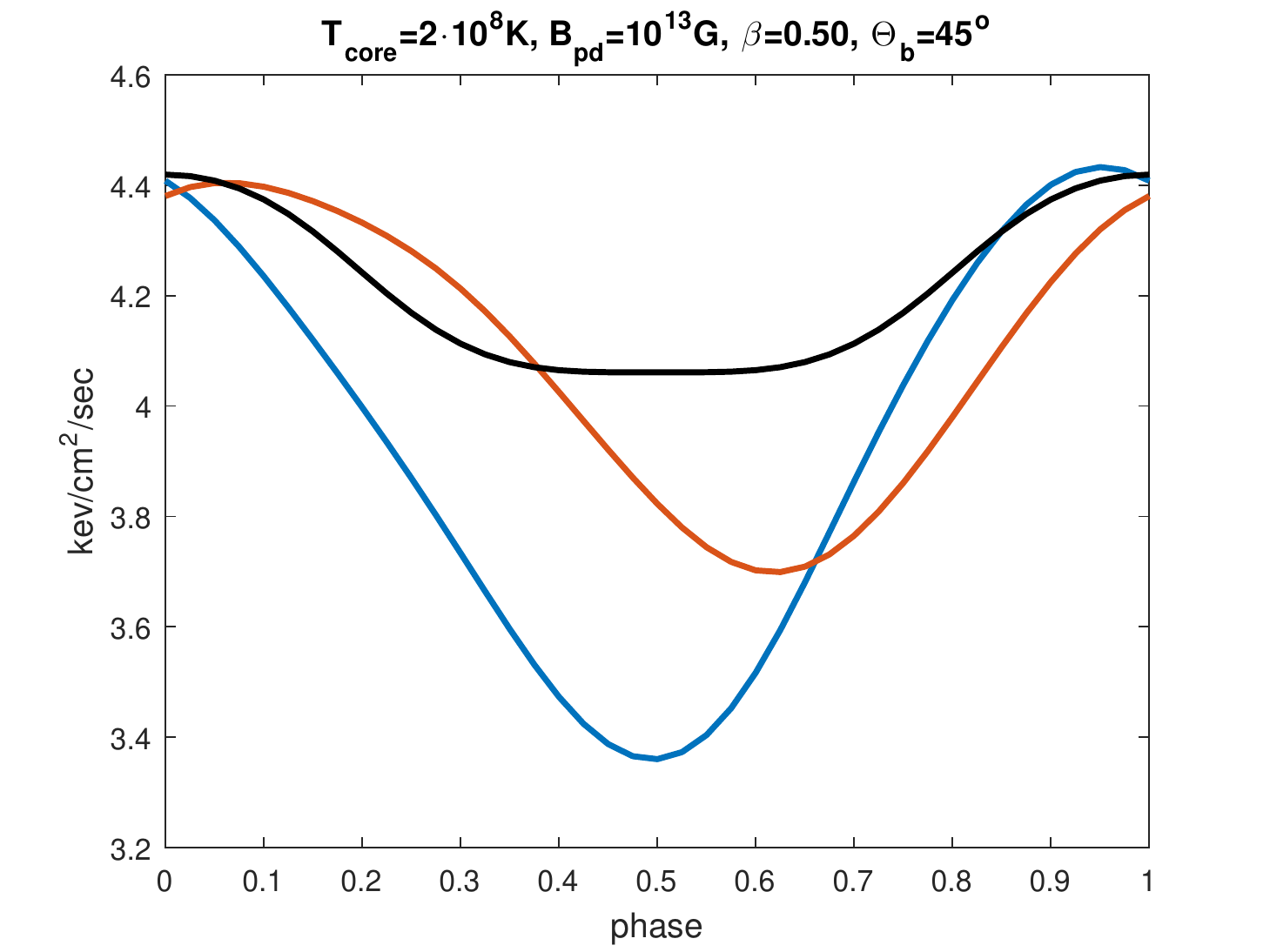}
	\includegraphics[width=9.0cm,height=4.5cm]{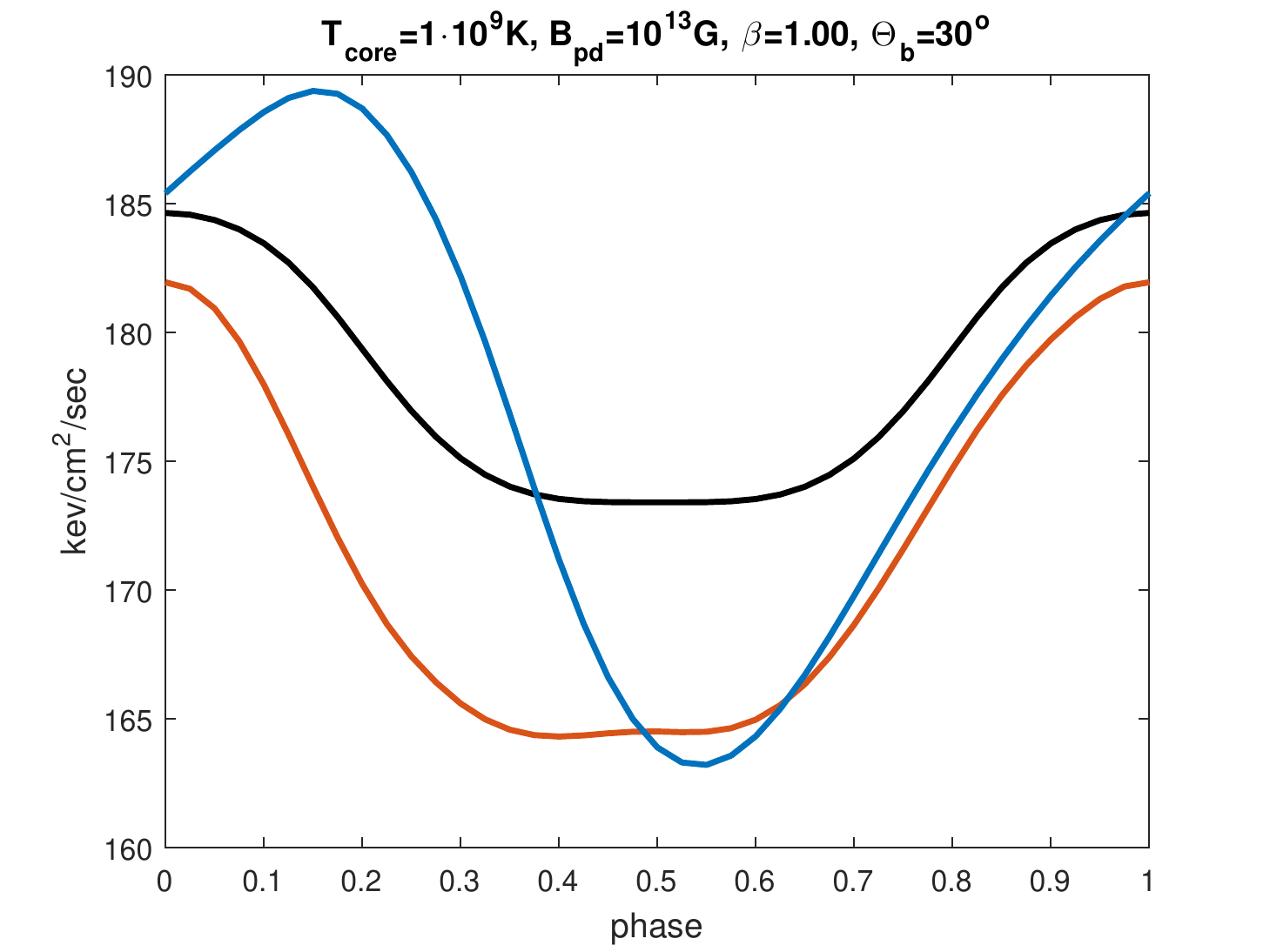}
	\caption{Light curves for the different $T_{core}, \beta$ and $\Theta_b$ for $\alpha_d = \zeta = \pi/4$ and $\alpha_q = \alpha_d$ (ones from Fig.\ref{fig:Tsurf1},\ref{fig:Tsurf23}). Blue lines correspond to the case, when only the $"$smaller$"$ pole, and red ones correspond to the $"$larger$"$ one. Black lines correspond to the light curves from the NS with a purely dipolar field.}
	\label{fig:lc1p}
\end{figure}

On the Fig.\ref{fig:lc1p} the light curves for the Case 2 are shown, where the blue lines correspond to the
smaller caps, while the red ones correspond to the larger caps. For both polar caps the pulsations are
amplified (see Table \ref{tab:table2} for the $PF$) in comparison to the NS with the purely dipolar field.
When the belt in the temperature distribution is the only one, one side of the pulse profile looks similar
to a straight line, when the smaller cap is visible, and the curve takes an irregular shape, when the larger
cap is observed. This linear dependence of an observed flux on the rotation phase makes the quadrupolar field
to be distinguishable. For the star with two cold belts the pulse minimum is shifted from the half period on
the light curve, if the energy has maxima on the boundaries of the curve picture (a blue line on the lowest
panel of Fig.\ref{fig:lc1p}),so that one slope on the light curve is more narrow, than the other one.
Such skewness in the pulse profile is inherent to RX J0806.4-4123 (see Fig.4 in \cite
{Haberl04}).

\begin{table}
	\caption{$PFs$ (\%) which correspond to the two-peaked light curves from the Fig.\ref{fig:lc1p} for
the dipolar field (black lines) and for the $"$larger$"$ (red lines) and $"$smaller$"$ (blue lines) hot caps.  }
	\label{tab:table2}
	\begin{center}	
		\begin{tabular}{| c| c|c|c|}	
			\hline
			  &  $\beta=0.75$ & $\beta=0.50$& $\beta=1.00$ \\
			    & $\Theta_b=60^o$ &  $\Theta_b=45^o$& $\Theta_b=30^o$ \\ \hline
			dipole & 5.1 & 4.2 & 3.1 \\
			$"$larger$"$ pole & 10.0 & 13.8 & 7.4 \\
			$"$smaller$"$ pole & 6.1 & 8.69 & 5.1 \\
			
			\hline
		\end{tabular}
	\end{center}
\end{table}	

\begin{figure}
	\centering
	\includegraphics[width=9.0cm,height=4.5cm]{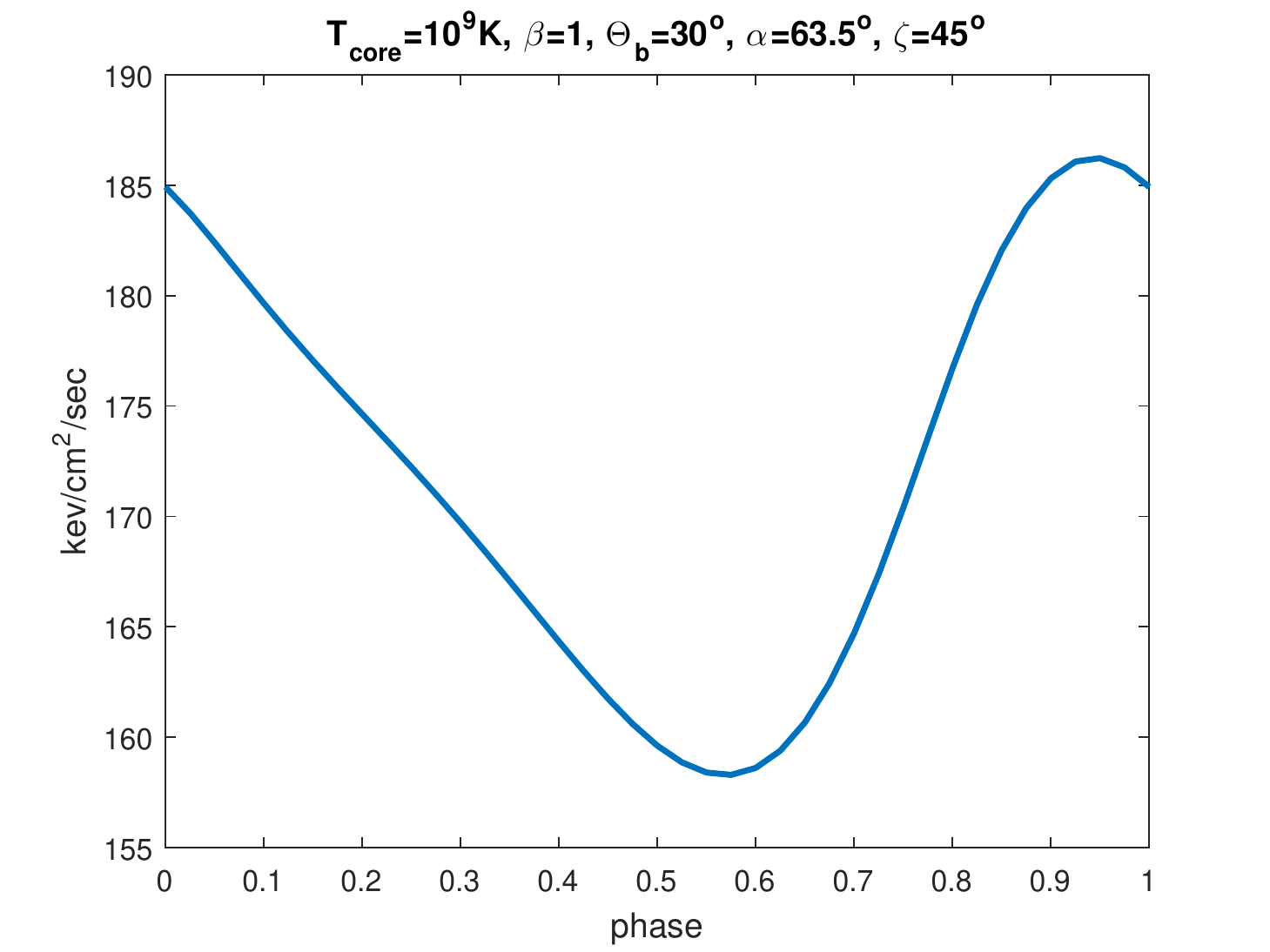}
	
	\caption{A light curve for the temperature distribution from the lower panel of Fig.\ref{fig:Tsurf23}. The viewing angle is $\zeta=45^o$, and $\alpha_d = 63.5^o$, $\alpha_q = 45^o$. }
	\label{fig:lc1psp}
\end{figure}

The latter situation shows itself more pronounced, when $\alpha_d\ne \zeta$. Although we consider only
$\alpha_d=\zeta$, this constraint may be artificial, and some types of light curves can be missed. For
example, we have calculated the light curve for the position of the dipolar component, which differs from
the viewing position (Fig.\ref{fig:lc1psp}). Such case may provide very non-symmetric pulse profile, about
$70\%$ of the flux dependence on the phase is described by a slope linear function. This feature on the
light curve may indicate an existence of the second belt, so it corresponds to the strong quadrupolar
component.

We do not affirm, that the synthesized light curves fit all the observed properties of the data from RX J0420 and RX J0806, because we have not taken into account, at least, absorption features. It was done properly by \cite{zaneturolla} using population analysis of models with coaxial dipole-plus-quadrupole fields. The calculated light curves for the NS with non-coincident magnetic axes of dipole and quadrupole can provide similar features as the listed sources. Our model also provides non-antipodal hot caps in the NS surface temperature distributions, which is possibly can be applied to RX J0720. To construct the curves, which fit the real data, the population analysis should be done like in (\cite{zaneturolla}). For 3D calculations it may be computationally expensive.

\section{Conclusions}

In this work we have studied the three-dimensional effects of the anisotropic heat transfer in outer layers
of the neutron star with the inclusion of the simplest core-configurations of dipole-plus-quadrupole magnetic
fields. We have self-consistently solved the stationary 3D heat transfer equation in the NS crust adopting
our model of the thermal structure of the outer envelope, which was built and discussed in the Paper 1, where
we had calculated axisymmetric heat transfer. We have obtained the temperature distributions in the NS
crust and on its surface, using the original numerical technique and analytically obtained tensorial electron
thermal conductivity coefficient by \cite{bisn2}. 
\newline
For the computed surface temperature distributions we have built thermal light curves using
a composite black-body model. The main purpose of this part of our study is to find some qualitative features on the synthetic light curves, which may indicate the non-coaxial field. Presence of the non-coaxial quadrupolar component may affect the light curves by strengthen the pulsations of
an observable flux and changing the shape of the pulse profile, as well as making it non-symmetric. Existence of a quadrupolar component in the magnetic field can be detectable, because the synthesized light
curves differ both from the pure-dipolar ones and ones with inclusion of the coaxial toroidal or crustal field
(\cite{heat2, heat3}), extending the $"$zoo$"$ of observational properties from thermally emitting NSs.

\section*{Acknowledgements}

This work was partially supported by the Russian Foundation for Basic Research (projects no. 18-02-00619, 18-29-21021 and 20-02-00455).


\section*{Data Availability}

The data underlying this article will be shared on reasonable request to the corresponding author.






\appendix

\section{Radiative opacities}
Radiative opacity is represented mostly by electron Thompson scattering, free-free and bound-free absorption.
In an absence of the magnetic field in non-relativistic limit Thompson opacity is given by \cite{BLP}:
\begin{equation}
K_{Th} = \frac{n_e\sigma_T}{\rho} = \frac{8\pi}{3}\big(\frac{e^2}{m_ec^2}\big)^2\frac{n_e}{\rho},
\label{opacity_Th}
\end{equation}
where $\sigma_T$ is a Thompson scattering cross-section, $\frac{e^2}{m_ec^2}$ is a classical electron radius.

A cross-section of a free-free absorption with taking into account spontaneous and stimulated emission
in non-relativistic limit in local thermodynamic equilibrium is given by:

\begin{equation}
\sigma^*_{aff} = \frac{4\pi}{3\sqrt{3}}\frac{Z^2e^6}{m_e^2chv\nu^3}g_{ff}\big(1 - e^{-h\nu/k_BT}\big),
\label{crosssection}
\end{equation}
where $v$ is an electron velocity, $\nu$ is a photon frequency, $g_{ff}$ is a Gaunt factor, which takes
into account quantum corrections to classical formula.
To obtain the absorption coefficient (\cite{bisn2001}) on one frequency, we have to average
\eqref{crosssection} with a Fermi-Dirac statistics:

\begin{displaymath}
\begin{gathered}
\alpha^{\nu}_{ff} = \frac{8\pi m_e^3}{Am_uh^3}\int_{0}^{\infty}\frac{\sigma^*_{aff} v^2dv}{1 +
\exp(\frac{mv^2}{2k_BT}-\chi)}q_{ff}, \\
q_{ff} = \bigg(1+\exp\bigg(\chi - \frac{h\nu}{k_BT}-\frac{mv^2}{2k_BT}\bigg)\bigg)^{-1},
\end{gathered}
\end{displaymath}
here $m_u$ is an atomic mass unit, and factor $q_{ff}$ determines a fraction of vacant electron states
in a degenerate gas. An integration gives:

\begin{equation}
\alpha^{\nu}_{ff} = \frac{4\pi}{3\sqrt{3}}\frac{8\pi
Z^2e^6}{Am_uch^4\nu^3}g_{ff}k_BT\log\bigg(\frac{e^{\chi}+1}{e^{\chi-h\nu/k_BT}+1}\bigg).
\label{aver_FD}
\end{equation}
To obtain the opacity expression for free-free transitions, it is necessary to derive the Rosseland meaning
(e.g. \cite{radiative}) from ~\eqref{aver_FD} in the following manner:

\begin{displaymath}
K_{ff} = \frac{\int_{0}^{\infty}\frac{1}{\alpha^{\nu}_{ff}}
\frac{dB_{\nu}}{dT}d\nu}{\int_{0}^{\infty}\frac{dB_{\nu}}{dT}d\nu}.
\end{displaymath}
In the formula above $B_{\nu}(T) = \frac{2h\nu^3}{c^2}\frac{1}{e^{h\nu/k_BT} - 1}$ is an intensity
of equilibrium Planck black-body radiation.
The Rosseland mean opacity for non-degenerate case reads (\cite{bisn2001})

\begin{equation}
K^{nd}_{ff} = 4.34\cdot 10^{22}\frac{\rho}{T^{7/2}}\frac{Z^2}{A} cm^2/g,
\label{ffnd}
\end{equation}
and in strongly degenerate limit the value of $K_{ff}$ is given by

\begin{equation}
K^{sd}_{ff} = 0.036\frac{32\pi^2}{3\sqrt{3}}\frac{e^6g_{ff}}{m_uchk_B^2}\frac{Z^2}{AT^2}.
\label{ffsd}
\end{equation}
A cross-section of bound-free absorption os given by the following formula in the non-relativistic limit
(e.g. \cite{bisn2001}):

\begin{equation}
\begin{gathered}
\alpha^{\nu}_{bf} = \frac{4}{3}\sqrt{\frac{2\pi}{3}}\frac{e^6h^2}{m_e^{3/2}cm_u(k_BT)^{7/2}}\frac{Z^2}{A}
n_e g_{bf}\times \\
\times\bigg[\frac{1}{n}\frac{E_b}{k_BT}\exp^{E_b/k_BT}\bigg(\frac{k_BT}{h\nu}\bigg)^3q_{bf}\bigg], \\
E_b = \frac{2\pi^2m_eZ^2e^4}{h^2n^2}, \\
q_{bf} = \bigg(1+\exp\bigg(\chi - \frac{h\nu}{k_BT}+\frac{E_b}{k_BT}\bigg)\bigg)^{-1},
\end{gathered}
\label{bfop}
\end{equation}
where $E_b$ is an energy value of an energy level of a bounded electron in a hydrogen-like ion, $n$ is an
energy level number, $g_{bf}$ is a Gaunt factor, and $q_{bf}$ is a degeneracy correction factor. To obtain
the opacity coefficient, we have to summarize an expression in square brackets of $\alpha_{bf}^{\nu}$
in \eqref{bfop} over all bound states and the make a Rosseland averaging of this expression. For
non-degenerate electrons we use an expression of $K_{bf}$ from the book by \cite{bisn2001}:

\begin{equation}
K_{bf}^{nd}= 7.23\cdot 10^{24}\frac{\rho}{T^{7/2}}\frac{Z^2}{A}\frac{g_{bf}}{t} cm^2/g,
\label{bfnd}
\end{equation}
where factor $\frac{t}{g_{bf}}$ takes values from 1 to 10. With increasing the density the electron gas in
the NS outer envelope goes fast to a strong degeneracy. To take into account an influence of the degeneracy
on the bound-free absorption, let us write a sum $\alpha_{bf}^{\nu}$ in \eqref{bfop} over the electron
bound states:

\begin{equation}
\alpha^{\nu}_{bf} = K_0\sum_{n=1}^\infty \frac{E_b}{nk_BT}\frac{e^{-\chi}}{x^3}\frac{e^x-1}{e^{x-\frac{E_b}{k_BT}-\chi} + 1},
\label{bfsd1}
\end{equation}
where $K_0$ = $K_{bf}^{nd}\cdot t$, and $x = \frac{h\nu}{k_BT}$. In the expression above the chemical
potential $\chi \gg 1$, and mostly only very hard photons in a $"$tail$"$ of a Planck spectrum are
absorbed effectively, and they will not contribute sufficiently in the mean opacity. Neglecting an
exponent $e^{x-\frac{E_b}{k_BT}-\chi}$ in \eqref{bfsd1} and averaging over $n$ and frequencies according
to \cite{MarShv} for the non-degenerate case, the following expression can be obtained for the degenerate case:

\begin{equation}
K_{bf}^{sd} = K_0 e^{-\chi} \frac{g'_{bf}}{t'},
\label{bfsd}
\end{equation}
factor $\frac{t'}{g'_{bf}}$ also takes values from 1 to 10. It is seen, that for the case of the
degenerate electron gas bf-opacity decreases exponentially with the growth of the density.

To use formulae \eqref{ffnd},\eqref{ffsd} for the ff-transitions and \eqref{bfnd},\eqref{bfsd} for
the bf-transitions in our calculations, we have to stitch them continuously, for example, in the following way:

\begin{equation}
K_{ff,bf} = K_{ff,bf}^{nd} \frac{1}{1 + e^{m\chi}} + K_{ff,bf}^{sd} \frac{e^{m\chi}}{\varepsilon + e^{m\chi}},
\label{Kbf}
\end{equation}
where $m,\varepsilon>1$ are numbers, which determine the smoothness of transition from one limit to another.
We note, that in $K_{bf}^{sd}$ it is necessary to replace $\chi = \frac{\mu_e}{k_BT}$ by its absolute value.
Thus, in the absence of magnetic field the value of the radiative opacity is composed of Thompson, free-bound
and free-free ones: $K_r(\rho,T)_{B=0} = K_{Th} + K_{bf} + K_{ff}$.

In presence of the strong magnetic field the photon opacity is reduced and becomes anisotropic. We have
taken into account an effect of the magnetic field on the opacity in the same manner, as in \cite{PY01}.
Authors of that work have built an analytical approximation of numerically obtained magnetic correction
factors from \cite{sya1980} for Thompson scattering and free-free absorption. The influence of magnetic
field on the bound-free opacity is assumed to be the same as on the free-free one.


\section{Numerical implementation}
\subsubsection*{Operator formulation of the problem}
With an approach suggested and discussed in \cite{ard1}, we have to include boundary conditions in
an operator-difference form of a considered problem. Let us write the system of equations \eqref{heattr}
in a difference form in a whole region using a cell-node approximation:

\begin{equation}
\begin{cases}
\nabla_{\times}\cdot \kappa^n\cdot\nabla_{\bigtriangleup}T^n + \delta_1\Phi \cdot \kappa^n \cdot\Phi T^n + \\
+ \delta_2\Phi\kappa^n\cdot\Phi T^n = 0, \\
\delta_1T^n = T_{core}, \\
\delta_2(\kappa^n\cdot\Phi T^n + \vec{n}\sigma T_s^4)= 0,
\end{cases}
\label{heattr_num}
\end{equation}
where the notations are adopted from (\cite{ard1}): $\nabla_\times\cdot$ and $\nabla_\Delta$ are difference
approximations of differential divergence and gradient operators, $\Phi$ is a boundary operator, it corresponds
to a derivation procedure on the boundary.
Here operator $\delta$ is defined as follows: it equals to $0$ in the interior mesh nodes and to $1$ on the
boundary surface, indexes 1 and 2 correspond the sort of a particular bound, 1 - inner and 2 - outer bound.
Index $n$ correspond to an iteration number in $"$time$"$, and $T_s^n = T_s^n(\delta_2T^{n-1})$. Temperature
is defined in the nodes, and magnetic field and density are defined in the cells and boundary nodes of
the mesh.

After acting on \eqref{heattr_num} with scalar boundary operator and subtracting it from the first equation,
we obtain the following system:

\begin{displaymath}
\begin{cases}
\nabla_{\times}\cdot\kappa^n\cdot\nabla_{\bigtriangleup}T^n + \delta_1\Phi \cdot \kappa^n \cdot\Phi T^n - \delta_2\Phi\cdot\vec{n}\sigma T_s^4 = 0 \\
\delta_1T^n = T_{core}
\end{cases}
\end{displaymath}
After that let us allocate the first (inner) boundary in the first equation from \eqref{heattr_num} and
multiply the first equation from \eqref{heattr_num} by $\delta_1$, and after substituting one equation
from another and a substitution the Dirichlet boundary condition on the inner boundary, we can derive the
final operator-difference equation:

\begin{equation}
\begin{gathered}
(I - \delta_1)\nabla_{\times}\cdot \kappa^n\cdot\nabla_{\bigtriangleup}(I - \delta_1){T^n} + \\
+ (I - \delta_1)\nabla_{\times}\cdot \kappa^n\cdot\nabla_{\bigtriangleup}T_{core} -
\delta_2\Phi\cdot\vec{n}\sigma T_s^4 = 0
\end{gathered}
\label{final_HT}
\end{equation}
where $I$ is a unit operator. A resulting operator equation is a finite-difference approximation of
the considered boundary problem \eqref{heattr}.

\subsubsection*{Algorithm of solving a heat transfer equation in a NS crust}

In this work we look for the stationary solution for the boundary problem \eqref{heattr}. This problem has
to be solved self-consistently, because surface temperature $T_s$ in the outer boundary condition is a
function of a temperature in the crust itself. We implemented an iterative procedure of relaxation: the
problem is solved with boundary conditions of first type on the inner boundary and of the second type on
the outer one $n$ times until inequality $max\big|\frac{T_s^n-T_s^{n-1}}{T_s^{n-1}}\big|<\epsilon$ is
not satisfied, where $n$ is an iteration number, $\epsilon$ is a some small number. After each iteration
the value of $T_s$ is specified with the $T_s-T_b$-relationship with the surface temperature distribution,
obtained from the previous iteration. The value $T_s^0$ for the boundary condition on the first iteration
is obtained from the initial approach of the crust temperature. In some sense, this procedure is equivalent
to a solving a time-dependent heat transfer equation with boundary conditions of the first and the third
types, while the value of the heat flux density  $\sigma T_s^4(T)$ in the outer boundary condition
from \eqref{heattr} is taken from the previous $"$time$"$-step.

On each $"$time$"$-step the system \eqref{heattr} is solved with the Basic operators method, described
in \cite{kondmoisBKGl}.

The operator-difference equation \eqref{final_HT} is nonlinear in $T^n$ and should be solved with Newton
method of solving systems of nonlinear equations, and appeared system of linear algebraic equations on the
each Newton method iteration is solved by the iterative Seidel method.

Thus, following the procedure described above, the self-consistent temperature distribution can be found
in a crust volume and on the NS surface.

\label{lastpage}
\end{document}